\documentclass[twocolumn,amssymb, nofootinbib, nobibnotes, aps, prd, superscriptaddress]{revtex4-2}
\usepackage{amsmath}
\usepackage{amssymb}
\usepackage{booktabs}
\usepackage{etoolbox}
\usepackage[T1]{fontenc}
\usepackage{graphicx}
\usepackage{hyperref}
\usepackage{siunitx}
\usepackage{xcolor}

\newcommand{\bsh}{\texttt{BinarySkyHough }}
\newcommand{\sh}{\texttt{SkyHough }}
\newcommand{\Tsft}{T_{\text{SFT}}}
\newcommand{\abs}[1]{\left|#1\right|}
\newcommand{\fsemi}{2\hat{\mathcal{F}}}

\graphicspath{{figures/}}

\newtoggle{fullauthorlist}
\togglefalse{fullauthorlist}

\newtoggle{endauthorlist}
\toggletrue{endauthorlist}

\begin{document}

\title{All-sky search in early O3 LIGO data for continuous gravitational-wave 
signals from unknown neutron stars in binary systems}

\iftoggle{endauthorlist}{
 %
 %
 \let\mymaketitle\maketitle
 \let\myauthor\author
 \let\myaffiliation\affiliation
 \author{The LIGO Scientific Collaboration and the Virgo Collaboration}
}{
 %
 %
 \iftoggle{fullauthorlist}{


\author{R.~Abbott}
\affiliation{LIGO Laboratory, California Institute of Technology, Pasadena, CA 91125, USA}
\author{T.~D.~Abbott}
\affiliation{Louisiana State University, Baton Rouge, LA 70803, USA}
\author{S.~Abraham}
\affiliation{Inter-University Centre for Astronomy and Astrophysics, Pune 411007, India}
\author{F.~Acernese}
\affiliation{Dipartimento di Farmacia, Universit\`a di Salerno, I-84084 Fisciano, Salerno, Italy  }
\affiliation{INFN, Sezione di Napoli, Complesso Universitario di Monte S.Angelo, I-80126 Napoli, Italy  }
\author{K.~Ackley}
\affiliation{OzGrav, School of Physics \& Astronomy, Monash University, Clayton 3800, Victoria, Australia}
\author{A.~Adams}
\affiliation{Christopher Newport University, Newport News, VA 23606, USA}
\author{C.~Adams}
\affiliation{LIGO Livingston Observatory, Livingston, LA 70754, USA}
\author{R.~X.~Adhikari}
\affiliation{LIGO Laboratory, California Institute of Technology, Pasadena, CA 91125, USA}
\author{V.~B.~Adya}
\affiliation{OzGrav, Australian National University, Canberra, Australian Capital Territory 0200, Australia}
\author{C.~Affeldt}
\affiliation{Max Planck Institute for Gravitational Physics (Albert Einstein Institute), D-30167 Hannover, Germany}
\affiliation{Leibniz Universit\"at Hannover, D-30167 Hannover, Germany}
% ...

\collaboration{The LIGO Scientific Collaboration and the Virgo Collaboration}



 }{
 \author{The LIGO Scientific Collaboration and the Virgo Collaboration}
}
}

\date{\today}

\begin{abstract}

    Rapidly spinning neutron stars are promising sources of continuous gravitational waves.
    Detecting such a signal would allow probing of the physical properties of matter under extreme conditions.
    A significant fraction of the known pulsar population belongs to binary systems. Searching for unknown 
    neutron stars in binary systems requires specialized algorithms to address unknown orbital frequency 
    modulations.
    We present a search for continuous gravitational waves emitted by neutron stars in binary systems  
    in early data from the third observing run of the Advanced LIGO and Advanced Virgo detectors using the 
    semicoherent, GPU-accelerated, \bsh pipeline. The search analyzes the most sensitive frequency band of 
    the LIGO detectors, $50 - 300\;\textrm{Hz}$. Binary orbital parameters are split into four regions,
    comprising orbital periods of $3 - 45\;\text{days}$ and projected semimajor axes of $2 - 40$ light-seconds.
    No detections are reported. We estimate the sensitivity of the search using simulated continuous wave signals, 
    achieving the most sensitive results to date across the analyzed parameter space.

\end{abstract}

\maketitle

\section{Introduction}

Continuous gravitational waves (CWs) are a long-lasting form of gravitational radiation.
For ground-based interferometric detectors the canonical sources are rapidly spinning neutron stars 
(NSs) sustaining a quadrupolar deformation. Several emission mechanisms have been proposed, such as 
crustal deformations, r-modes, or free precession  (see \cite{Sieniawska_2019} for a recent review). 
Detecting CWs would probe the physics of such compact objects, leading us 
to a better understanding of the equation of state of matter under extreme conditions. 
More exotic types of CW sources are also theorized, such as boson clouds around spinning black holes 
\cite{PhysRevD.83.044026}. 

Every CW search method assumes certain information about the intended sources.
All-sky searches, such as the one reported in this paper, impose the least constraints on the CW
emission. The latest results obtained by the LIGO-Virgo collaboration using Advanced LIGO \cite{AdvancedLIGO}
and Advanced Virgo \cite{AdvancedVirgo} data, covering targeted (known pulsars), directed (known sky
locations), and all-sky searches, can be found in \cite{PhysRevD.99.122002, Abbott_2019, PhysRevD.100.122002, 
IsolatedO2, Abbott:2020lqk, J0537}.

All-sky searches require highly efficient analysis methods because they must account for a Doppler modulation due 
to the Earth's movement with respect to the Solar System Barycenter (SSB), an effect that depends on sky 
position. In principle, one can construct a search pipeline using fully coherent matched filtering; for wide parameter
space searches, however, such an approach quickly becomes computationally unaffordable \cite{PhysRevD.57.2101}.
As a result, semicoherent methods are used, splitting the data stream into smaller time segments that can 
be coherently analyzed. Then, per-segment results are combined according to the expected
frequency evolution of the template under analysis. This method reduces the computational cost of a search 
while achieving a reasonable sensitivity.

Only a small fraction of the expected population of galactic NSs has been detected electromagnetically 
\cite{Rajwade:2018xxf}. Through gravitational waves we could access these unknown populations of NSs.
About half of the NSs detected using electromagnetic means within the most sensitive frequency band of current 
ground-based detectors are part of a binary system \cite{BSH, atnfPaper}. Searches for CWs from this class of 
NSs pose an additional, substantial computational challenge compared to standard all-sky searches that target 
isolated NSs because additional unknown binary orbital parameters increase the search parameter space 
dimensionality. As a result, one must use specialized methods in order to search for this type of signal.

We present an all-sky search for CWs produced by NSs in binary systems using the semicoherent \bsh pipeline 
\cite{BSH}. It builds upon \sh \cite{KrishnanSintes}, inheriting its characteristic noise robustness and 
computational efficiency, and uses Graphics Processing Units (GPUs) to speed up the core part of the search. 
The concept of \bsh is to compute search statistics over the parameter space and to use those statistics to rank
the interesting regions for subsequent follow up using more sensitive, computationally demanding techniques.
This balance between sensitivity and computational cost has proven effective in previous searches of the LIGO O2
observing run data using both the isolated \sh \cite{IsolatedO2} and \bsh \cite{BSHO2} flavors of this pipeline.

In Sec.~\ref{sec:signal_model} we introduce the signal model;
Sec.~\ref{sec:sfdb} describes the early third observing run of the Advanced LIGO and Advanced Virgo detectors;
Sec.~\ref{sec:bsh} briefly describes the main analysis pipeline; Sec.~\ref{sec:pp} introduces the first
post processing stage; and in Sec.~\ref{sec:ul} we estimate the sensitivity of this search.
In Sec.~\ref{sec:fu} we further analyze the most significant outliers and rule them out as non-astrophysical
candidates. We present our conclusions in Sec.~\ref{sec:con}.

\section{Signal Model}\label{sec:signal_model}

A non-axisymmetric neutron star spinning about one of its principal axes is expected to emit gravitational
waves at twice its rotation frequency $f_{0} = 2 f_{\textrm{rot}}$ with a strain amplitude given by 
\cite{JK1999}
\begin{equation}
\begin{split}
    h(t) = & h_{0} \; [ F_{+}(t;\psi, \hat{n}) \; \frac{1 + \cos\iota}{2} \;\cos \phi(t) \\
    & + F_{\times}(t; \psi, \hat{n}) \; \cos \iota  \;\sin\phi(t) ] \;,
\end{split}
\end{equation}
where $F_{+,\times}$ are the antenna patterns of the interferometric detectors, depending on the polarization
angle $\psi$ and the sky position $\hat{n}$ of the source; $h_{0}$ and $\cos\iota$ are the 
characteristic CW amplitude and the cosine of the inclination of the source with respect to the line of sight, 
respectively; $\phi(t)$ represents the phase of the gravitational wave signal.

The CW amplitude $h_0$ can be expressed in terms of the physical properties of the source once an emission
mechanism has been assumed. The three principal moments of inertia of a non-axisymmetric NS are given by
$I_{\textrm{x}},\; I_{\textrm{y}},\; I_{\textrm{z}}$, and the equatorial ellipticity is given by
$\epsilon = \abs{I_{\textrm{x}} - I_{\textrm{y}}} / I_{\textrm{z}}$, assuming the spin axis is aligned with
$I_{\textrm{z}}$. The gravitational wave amplitude can be expressed as
\begin{equation}
    h_{0} = \frac{4 \pi^2 G}{c^4} \frac{I_{\textrm{z}}\epsilon}{d} f_{0}^2 \; ,
    \label{eq:h0}
\end{equation}
where $d$ denotes the distance to the source from the detector, $f_{0}$ the gravitational wave frequency, and
$G$ and $c$ respectively refer to the gravitational constant and the speed of light. We can further relate 
this quantity to the mass quadrupole $Q_{22}$ of the star through the equatorial ellipticity
\begin{equation}
    \epsilon = \sqrt{\frac{8 \pi}{15}} \frac{Q_{22}}{I_{\textrm{z}}}\;.
\end{equation}

We can describe the signal phase via Taylor expansion with respect to a fiducial starting
time $\tau_{0}$ in the source frame
\begin{equation}
    \phi(\tau) = \phi_0 + 2 \pi \left[ f_{0} \cdot \left(\tau - \tau_{0}\right) + \dots \right]\;,
    \label{eq:phase}
\end{equation}
where $\tau$ is the proper source frame time and $\phi_0$ represents the initial phase at $\tau_{0}$.
The number of higher order terms to include in this expansion depends on the population of NSs
under consideration.
After analyzing the ATNF pulsar catalog \cite{atnfPaper}, it was argued in \cite{BSH} that searching for 
NSs in binary systems need not take into account any spindown parameters when using datasets 
lasting for less than a few years.  As we will discuss in Sec.~\ref{sec:con}, this search remains sensitive 
to signals up to a certain spindown value, but there is an implicit limit on the astrophysical reach.

Because of the relative motion of the detector around the SSB and the relative motion
of the source around the Binary System Barycenter (BSB), the phase as measured by the detector at time 
$t$ is Doppler-modulated according to the timing relation
\begin{equation}
    \tau + a_p \sin\left[\Omega \left(\tau - \tau_{\textrm{asc}}\right)\right] = 
    t + \frac{\vec{r}(t) \cdot \hat{n}}{c} - \frac{d}{c}\;,
    \label{eq:timing}
\end{equation}
where $a_p$ represents the semi-major axis of the binary orbit projected onto the line of sight (measured in
light-seconds), $\Omega$ represents the orbital frequency of the source, $\tau_{\textrm{asc}}$ represents
the time of passage through the ascending node as measured from the source frame and $\vec{r}$ represents the
position of the detector in the SSB. In order to derive this expression, we assumed circular, Keplerian orbits;
the search remains sensitive, however, to signals from sources in binary systems up to a certain eccentricity 
as discussed in Section \ref{sec:priors} and \cite{BSH}.

We define a \emph{template} as $\lambda = \left\{f_{0}, \hat{n}, a_p, \Omega, t_{\textrm{asc}}\right\}$. 
The parameter space (i.e. the set of all templates searched) will be denoted as 
$\mathbb{P}$. The orbital period is related to the orbital angular frequency by $P = 2\pi / \Omega$.

We refer the reader to \cite{PhysRevD.91.102003} for a complete derivation of Eq.~\eqref{eq:timing} and a 
discussion about how to express Eq.~\eqref{eq:phase} in the detector frame. The gravitational wave 
frequency evolution associated to a template $\lambda$ as measured from the detector frame is thus
\begin{equation}
    f_{\lambda}(t) = f_{0} \cdot \left(1 + \frac{\vec{v}(t) \cdot \hat{n}}{c} 
    - a_{p} \Omega \cos\left[\Omega (t-t_{\textrm{asc}})\right] \right)\;,
    \label{eq:track}
\end{equation}
where $\vec{v}(t)$ refers to the detector velocity and $t_{\textrm{asc}}$ is akin to $\tau_{\textrm{asc}}$ 
measured from the detector frame. We choose the initial phase $t_{\textrm{asc}}$ to be located within the range 
$\left[t_{\textrm{mid}} - \frac{P}{2},\; t_{\textrm{mid}} + \frac{P}{2}\right]$, where
$t_{\textrm{mid}}$ represents the mean time between the start and the end of the run measured in GPS seconds.

\section{Data used}\label{sec:sfdb}
    
The first part of the third observing run of the Advanced LIGO and Advanced Virgo detectors (O3a) 
comprises six months of data collected from the 1st of April 2019 at 15:00 UTC to the 1st of October 
2019 at 15:00 UTC. Data was taken by the Advanced LIGO detectors, located in Hanford (Washington, USA, 
designated H1) and Livingston (Lousiana, USA, designated L1), together with the Advanced Virgo detector, 
located in Cascina (Pisa, Italy). We did not make use of Advanced Virgo data because of an unfavorable trade-off
between computing cost and expected sensitivity improvement of the search. The detector duty factor 
(the fraction of the run when the detector is collecting observational-quality data) was 71.2\% for H1 
and 75.8\% for L1. The implementation of instrumental upgrades has allowed the detectors to improve their 
overall sensitivities with respect to the previous observing run (O2) \cite{O3aRun}.

For the duration of the run, several artificial signals were injected into both detectors in order to calibrate 
and monitor their performance.
\emph{Calibration lines} are artificial monochromatic signals, injected at different frequencies in each 
detector to avoid coherent artifacts. They are used to monitor time-varying detector operating parameters.
\emph{Hardware injections}, on the other hand, are artificial quasi-monochromatic signals consistently injected 
into both detectors in order to mimic the effects of an actual CW signal present in both detectors. They are 
used to verify expected detector response and characterize calibrated data \cite{Sun_2020}.
Both of these artificial signals may interfere with CW searches in general, showing up as significant candidates 
due to their high strength in the detector spectrum.
Spectral artifacts in detector data can be produced by environmental or instrumental noise and also interfere
with CW searches \cite{PhysRevD.97.082002}.

The search was performed using Short Fourier Transforms (SFTs) created from the C00 
(initial calibration version) time-domain observing-quality strain data \cite{Viets:2017yvy}.
These SFTs were extracted from SFDB (Short Fourier Data Base) data \cite{Astone_2005}, which incorporates 
a time-domain cleaning procedure to avoid noise-floor degradation due to glitches and other forms of 
transient noise. Every SFT lies completely within \mbox{observing-quality} data.
Fourier transforms were computed using a Tukey-windowed baseline of \mbox{$\Tsft=\SI{1024}{\second}$} with
tapering parameter \mbox{$\beta_{\text{Tukey}} = 0.5$} and a $50\%$ overlap. These values are collected in
table \ref{table:misc}.

Following the same procedure used in the O2 \sh search \cite{IsolatedO2}, SFT data are split into two
datasets to be used in two different stages of the search. The first dataset, which we refer to as 
\emph{non-overlapping}, leaves out overlapping SFTs (i.e. every SFT starts at the end of the previous one).
The second dataset, which we refer to as \emph{overlapping}, contains all of the SFTs. 
Using the non-overlapping set for the first stage of the analysis reduces the computational cost of the 
search at a manageable loss in sensitivity. Table~\ref{table:sft} lists the number of SFTs in each of the
datasets. Datasets contain SFTs from \emph{both} LIGO detectors (i.e. we perform a \emph{multi-detector} search
\cite{BSH}).

\begin{table}[]
\begin{tabular}{ll}
    \toprule
    Search setup parameter & Value \\
    \midrule
    $\Tsft$ & $1024~\textrm{s}$ \\
    $\beta_{\text{Tukey}}$ & 0.5\\
    $T_{\textrm{obs}}$ & $14832675~\textrm{s}$\\
    $t_{\textrm{mid}}$ & $1245582821.5~\textrm{s}$\\
    $\Delta f$ & $0.125~\textrm{Hz}$\\
    \bottomrule
\end{tabular}
    \caption{Miscellaneous parameters used in the search. 
    $\Tsft$ denotes the time span employed to compute Short Fourier Transforms (SFTs). 
    $\beta_{\text{Tukey}}$ refers to the tapering parameter of the
    Tukey window, denoting the fractional length of the window's central unitary plateau.  
    $T_{\textrm{obs}}$ is the observing time of the run. $t_{\textrm{mid}}$ represents the
    mean time between the start and the end of the run measured in GPS seconds.
    $\Delta f$ refers to the bandwidth of the individual sub-bands analyzed by each computing job.}
    \label{table:misc}
\end{table}

\section{The Search Pipeline}\label{sec:bsh}

\begin{figure*}
    \includegraphics[width=\textwidth]{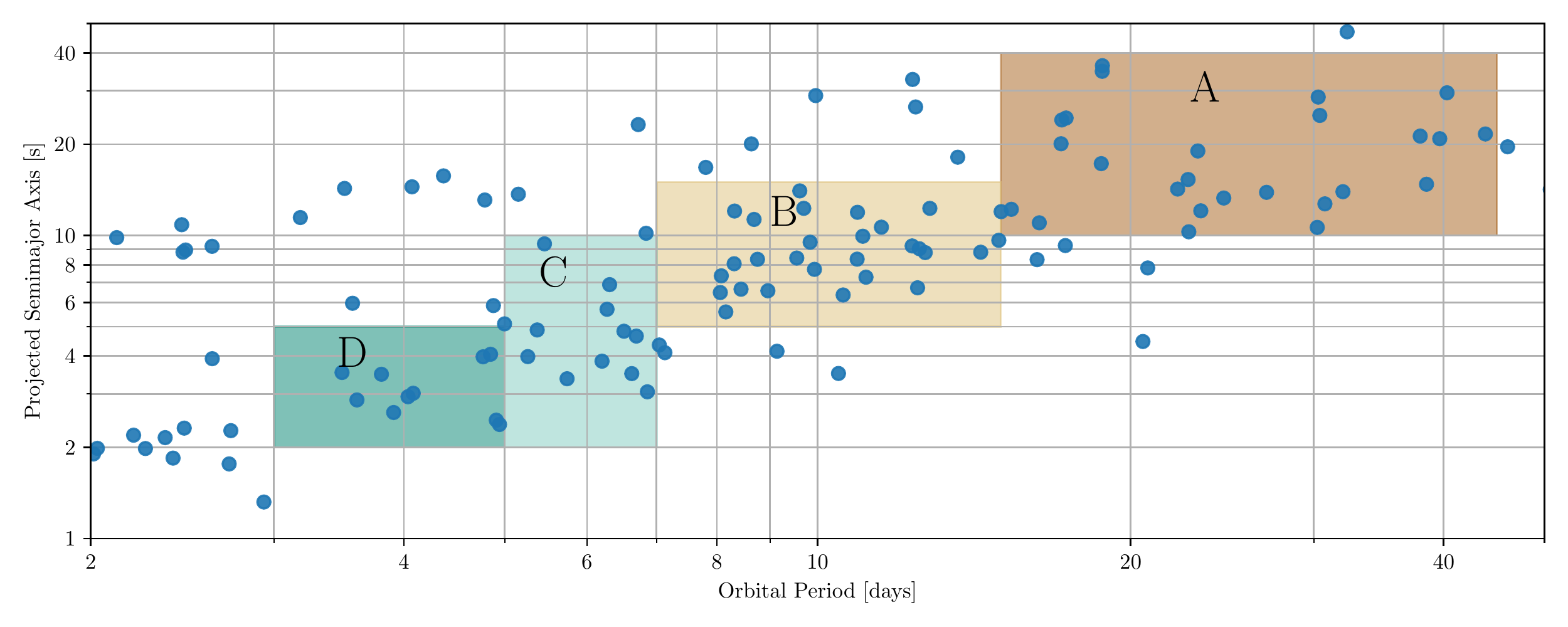}
    \caption{Binary orbital parameters considered by the present search. Solid color regions 
    denote parameter space regions in which a search was performed; blue dots mark binary
    orbital parameters corresponding to the known binary pulsar population. 
    Regions A, B, C and D were covered by the low-frequency analysis, while region B was 
    covered by the high-frequency analysis as well. Time of ascending node passage is taken 
    into account according to the orbital frequency, as explained in Sec.~\ref{sec:signal_model}.
    Pulsar population data was taken from \cite{atnfPaper} using \cite{psrqpy}.}
    \label{fig:binary_parameters}
\end{figure*}

We split the search into two main frequency bands: the low-frequency band, from $50~\textrm{Hz}$ to
$100~\textrm{Hz}$, and the high-frequency band, from $100~\textrm{Hz}$ to $300~\textrm{Hz}$. These
bands are further divided into $\Delta f =0.125\;\text{Hz}$ sub-bands, which constitute the basic 
working unit of our setup: each \emph{computing job} performs an all-sky search over one such sub-band, 
searching for binary modulated signals within a certain region of the binary parameter space among the ones
specified in Fig. \ref{fig:binary_parameters} and Table~\ref{table:binary_regions}. Because of the limited computing
power available, the high-frequency search focuses on a single binary parameter space region, denoted
as B in Table~\ref{table:binary_regions}; the low-frequency search is performed in all four binary parameter
space regions.

The search parameter space is gridded with templates as described in \cite{BSH}:
\begin{equation}
    \begin{split}
        \delta f_{0} &=\frac{1}{\Tsft}\;, \quad
        \delta \theta = \frac{ c/v }{\Tsft \; f_0 \; P_\mathrm{f}}\;,\quad
        \delta a_p = \frac{ \sqrt{6 m}}{\pi \Tsft f_0 \Omega}\;,\quad \\
        & \delta \Omega = \frac{\sqrt{72 m}}{\pi \Tsft f_0 a_p \Omega T_{\textrm{obs}}}\;,\quad
        \delta t_{\textrm{asc}} = \frac{\sqrt{6 m}}{\pi \Tsft f_0 a_p \Omega^2 }\:,\\
    \end{split}\
    \label{eq:resolution}
\end{equation}
where $\delta \theta$ refers to the angular sky position resolution, $v = \abs{\vec{v}}$ and $v/c \sim 10^{-4}$.
$T_{\textrm{obs}}$ denotes the observing time of the search, quoted in Table~\ref{table:misc}.
The variables $P_\mathrm{f}$ and $m$ are the so-called \emph{pixel factor} and \emph{mismatch} parameters,
which can be used to manually control the parameter space template density. 
In this search, we tune them in order to adjust the computing cost as we reach higher
frequencies, where template spacing naturally becomes finer. Table~\ref{table:frequency_regions}
summarizes the choices made for each of the frequency bands.

The pipeline uses the Hough transform to relate tracks in the digitized
spectrogram, as explained below, to points in the parameter space. For each point 
in the parameter
space $\lambda \in \mathbb{P}$ there is a corresponding track [see Eq. \eqref{eq:track}] 
of the time-frequency evolution, which denotes the instantaneous frequency of the
signal as observed by the detector. 

\begin{table}[]
\begin{tabular}{lcc}
    \toprule
    & Non-overlapping & Overlapping \\
    \midrule
    H1 & 10172 & 20577 \\
    L1 & 10962 & 22049 \\
    Total & 21234 & 42626 \\
    \bottomrule
\end{tabular}
    \caption{Number of Short Fourier Transforms (SFTs) in each of the datasets. Characteristics of these
    SFTs are summarized in table \ref{table:misc} and section \ref{sec:sfdb}.}
    \label{table:sft}
\end{table}
\subsection{Ranking statistics}

Let us assume the data can be described as a noise background plus a CW signal
\begin{equation}
x(t;\lambda) = n(t) + h(t; \lambda)\;.
\end{equation}
We start by computing the \emph{normalized power} of SFT data 
\begin{equation}
    \rho^{\alpha}_k = \frac{ \abs{\tilde{x}^{\alpha}_k}^2 }{\langle \abs{\tilde{n}^{\alpha}_k}^2 \rangle}\; ,
    \label{eq:normpower}
\end{equation} 
where tildes represents a Fourier transformed quantity, $k$ indexes frequency bins,
$\alpha$ indexes SFTs and $\langle \cdot \rangle$ denotes a running median average using 101 
frequency bins, as explained in \cite{BSH}. Each SFT $\alpha$ can be related to a certain starting time
$t_{\alpha}$, effectively obtaining a spectrogram where each bin $(\alpha,k)$ 
corresponds to the normalized power $\rho^{\alpha}_{k}$ present at a certain frequency bin
$k$ in a certain SFT $\alpha$. Then, we impose a normalized power threshold 
$\rho_\text{th} = 1.6$ to digitize the spectrogram, obtaining a discrete spectrogram 
populated by ones and zeros. 

For each template, we follow the corresponding track and define the first ranking statistic,
the \emph{number count}, as the weighted sum of ones and zeroes
\begin{equation}
    n(\lambda) = \sum_{(\alpha, k) \in f_{\lambda}}
    w^{\alpha}_{k}\;\mathcal{H}\left(\rho_{k}^{\alpha} - \rho_{\textrm{th}}\right)\;,
    \label{eq:numbercount}
\end{equation}
where $\mathcal{H}$ denotes the Heaviside step function and the weights $w^{\alpha}_{k}$
account for varying noise floor and antenna response effects \cite{Sintes_2006}.

\begin{table}[]
\begin{tabular}{cll}
    \toprule
    Binary Region & $P\;$[days] & $a_p\;$[s] \\
    \midrule
    A & [15, 45] & [10, 40] \\
    B & [7, 15] & [5, 15] \\
    C & [5, 7] & [2, 10] \\
    D & [3, 5]& [2, 5] \\
    \bottomrule
\end{tabular}
    \caption{Binary parameter space regions analyzed by the search, corresponding to the four colored
    regions in Fig. \ref{fig:binary_parameters}. Time of passage through the ascending node 
    $t_{\textrm{asc}}$ is searched along the interval specified in Sec.~\ref{sec:signal_model}.}
    \label{table:binary_regions}
\end{table}

\begin{table}[]
    \begin{tabular}{cccc}
    \toprule
        Label & Frequency [Hz] & $m$ & $P_\mathrm{f}$\\
    \midrule
        L & [50, 100) & 0.4 & 1\\
        H1 & [100, 125) & 0.4 & 1\\
        H2 & [125, 150) & 0.6 & 1\\
        H3 & [150, 200) & 0.9 & 0.8\\
        H4 & [200, 250) & 1.6 & 0.75\\
        H5 & [250, 300) & 2 & 0.7\\
    \bottomrule
\end{tabular}
    \caption{Mismatch and pixel factor configurations for the different frequency bands of the search.
    L refers to the low-frequency band; H1-5 refer to each of the five sub-bands into which the high-frequency 
    band was partitioned: 1 and 2 span 25Hz each, while 3 to 5 span 50~Hz each.}
    \label{table:frequency_regions}
\end{table}

The number count statistic can be efficiently computed by means of the Look Up Table (LUT) approach described in
\cite{KrishnanSintes}. Incidentally, this strategy simplifies the cost by analyzing multiple sky positions
(called \emph{sky patches}) together. The approach applies the Doppler modulation used to analyze a particular
frequency bin to a neighborhood of frequency bins. The sensitivity loss introduced by this approximation
is later compensated by re-analyzing the most significant candidates using their exact time-frequency tracks
\cite{delaJordana:2010za}. 

The re-analysis uses the \emph{weighted normalized power} statistic,
\begin{equation}
    \rho(\lambda) = \sum_{(\alpha, k) \in f_{\lambda}}
    w^{\alpha}_{k}\;\rho_{k}^{\alpha}\;.
    \label{eq:wnormpower}
\end{equation}
Using this new ranking statistic instead of simply re-computing Eq.~\eqref{eq:numbercount} along the exact track 
yields a \mbox{$10-20\%$} improvement in detection efficiency for the toplists (ranking of the most significant
candidates) based on the number-count statistic used here and discussed below 
\cite{IsolatedO2}.

In order to further select candidates across different sky patches, we compute a \emph{significance} statistic
by normalizing Eq.~\eqref{eq:wnormpower} to the \emph{expected noise values} derived in \cite{KrishnanSintes}:
\begin{equation}
    s_{\rho}\left(\lambda\right) = \frac{\rho(\lambda) - \bar{\rho}}{\sigma_{\bar{\rho}}}\;,
   \label{eq:significance}
\end{equation}
where $\bar{\rho}$ and $\sigma_{\bar{\rho}}$ represent the expected value and standard deviation of weighted
normalized power in pure Gaussian noise. This statistic removes any dependency on the 
sky position of the source due to the weights, being well suited for comparisons across different sky patches.

\subsection{Toplist construction}

\begin{figure}
    \includegraphics[width=0.5\textwidth]{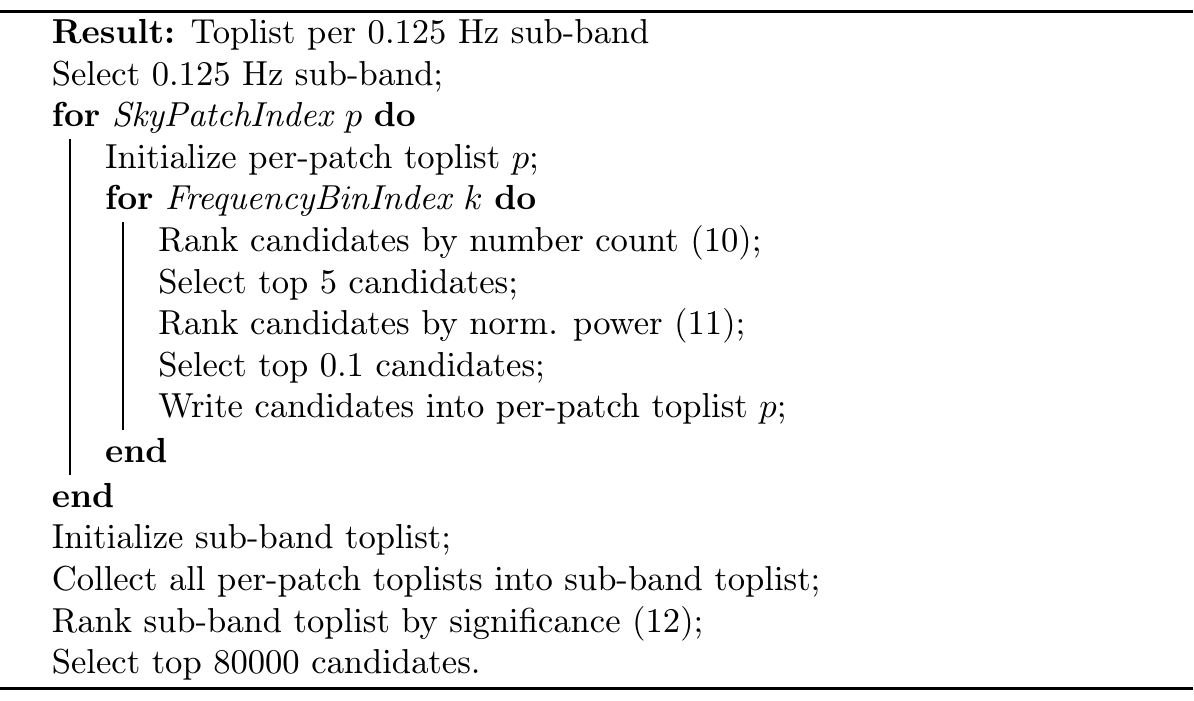}
    \caption{Explicit description of the \bsh toplist construction process.}
    \label{alg:toplist}
\end{figure}

Toplists are constructed frequency-bin wise across sky patches, as shown schematically in Fig. \ref{alg:toplist}.
For a given sky patch and frequency bin, the top 5\% of parameter space candidates are selected
according to the number count statistic Eq.~\eqref{eq:numbercount} using the LUT approach and the 
non-overlapping set of SFTs. Then, they are re-analyzed computing their corresponding normalized power 
Eq.~\eqref{eq:wnormpower} along the exact time-frequency evolution given by Eq.~\eqref{eq:track} using the 
overlapping set of SFTs. Finally, top candidates according to Eq.~\eqref{eq:significance} are collected into a
final toplist. We collect the top 80000 candidates from each $0.125~\textrm{Hz}$ sub-band.

This approach optimizes the GPU usage in the number count stage (preventing loud spectral 
artifacts from saturating the toplist) as each frequency bin provides a controlled number of candidates.

\section{Post Processing}\label{sec:pp}

Similar to previous searches \cite{IsolatedO2, BSHO2}, we apply a clustering algorithm to the resulting 
candidates in order to look for particularly interesting candidates. Clustering candidates reduces the total 
number of candidates to follow up since typically many candidates are found to be produced by a single source
(either a CW signal or an instance of instrumental noise).

We implement a new clustering algorithm using the frequency evolution of a candidate to define a 
parameter space distance \cite{Tenorio:2020cqm}. This choice allows the algorithm to naturally take into 
account the parameter space structure, avoiding the usage of ad hoc sky projections or mishandling
periodic boundary conditions.

After the cluster selection, we apply the well-known \emph{line veto}, used in previous searches 
(e.g. \cite{Abbott:2017mnu, Abbott:2018bwn, IsolatedO2, BSHO2}) in order to rule out non-astrophysical 
candidates.

\subsection{Clustering}

The clustering algorithm is summarized below; see \cite{Tenorio:2020cqm} for further details.
Given two candidates with template values $\lambda, \lambda^{*} \in \mathbb{P}$, we define the parameter space
distance as
\begin{equation}
    d(\lambda^{*}, \lambda) = \frac{\Tsft}{N_{\alpha}} \sum_{t_{\alpha}}
    \abs{f_{\lambda^{*}}(t_{\alpha}) - f_{\lambda}(t_{\alpha})}\;,
    \label{eq:tbd}
\end{equation}
where $f_{\lambda}(t_{\alpha})$ represents the instantaneous frequency of a CW produced by a source with 
parameters $\lambda$ as measured by the detector at time $t_{\alpha}$ and $N_{\alpha}$ denotes the number of
SFT timestamps used. Essentially, Eq.~\eqref{eq:tbd} is the average mismatch among time-frequency tracks. 

Clusters are formed by grouping together candidates from connected components,
i.e. each candidate in a cluster is closer than a maximum distance $d^{\textrm{th}}=1$
to at least one other candidate in the same cluster.
Final clusters are ranked according to the significance of their loudest candidate, which we will refer to 
as the \emph{cluster center}.

For each $0.125\;\textrm{Hz}$ toplist, the top 5 clusters according to their significance are selected. 
This leaves us with a total of 16000 clusters: 8000 for the high-frequency search and 2000 for each region 
of the low-frequency search.

\subsection{Line Veto}

\begin{table*}[]
    \begin{tabular}{rccccccccc}
    \toprule
    Regions & LA & LB & LC & LD & H1B & H2B &H3B & H4B & H5B  \\
        \midrule
        Initial Clusters & 2000 & 2000 & 2000 & 2000 & 1000 & 1000 & 2000 & 2000 & 2000 \\
        Vetoed by Identified Line & 366 & 359 & 359 & 373 & 44 & 0 & 32 & 30 & 30\\
    \midrule
        Surviving Clusters & 1634 & 1641 & 1641 & 1627 &  956 & 1000 & 1968 & 1970 & 1970 \\
        Fraction (\%) & 81.7 & 82.05 &  82.05 &  81.35 &  95.6  & 100  &  98.4  &  98.5  &
        98.5 \\
    \midrule
        Surviving Outliers after $\fsemi_{\textrm{th}}$ veto & 73 & 72 & 71 & 71 & 7 & 6 & 8 & 3 & 0\\
    \bottomrule
    \end{tabular}
    \caption{Numbers of cluster centers discarded by the line veto using lines present in \cite{IdentifiedLines}. 
    The number of surviving outliers after the follow-up stage (see Sec. \ref{sec:fu}) is here specified 
    for the sake of completeness. Five clusters were collected from each $0.125~\textrm{Hz}$ band: regions 
    H1B and H2B, being the only ones spanning $25~\textrm{Hz}$, yield a lower number of clusters.
    }
    \label{table:lines}
\end{table*}

Before the outlier follow up, we apply the line veto to the obtained cluster
centers. Using the list of identified narrow spectral artifacts \cite{IdentifiedLines}, the veto discards 
any candidate whose time-frequency track crosses an instrumental line, since such a candidate would 
likely become significant not because of astrophysical reasons but rather instrumental ones.

For every cluster center with parameters $\lambda$, we compute its 
\emph{bandwidth} 
\begin{equation}
    \mathcal{BW}(\lambda) = [\min_{t_{\alpha}} f_{\lambda}(t_{\alpha}),
    \max_{t_{\alpha}} f_{\lambda}(t_{\alpha})]\;.
\end{equation}
If the bandwidth of a candidate contains or overlaps with any of the lines present in 
\cite{IdentifiedLines}, then the candidate is discarded because of its likely non-astrophysical origin. 
This veto reduces the number of clustered candidates by $\sim\!20\%$ in the low-frequency search and by a few 
percent in the high-frequency search (see Table~\ref{table:lines}). This difference is to be expected, 
considering the greater amount of instrumental lines present at lower frequencies.

Other narrow spectral artifacts have not yet been identified as clearly non-astrophysical in origin in an 
unidentified list \cite{UnidentifiedLines}. Although this list has not been used to veto clustered candidates, 
some of them are consistent with artifacts in the unidentified list (see appendix \ref{appendix}).

\section{Sensitivity}\label{sec:ul}

The sensitivity of the search is determined using a similar procedure as for previous all-sky searches
\cite{Abbott:2017mnu, Abbott:2018bwn, IsolatedO2, BSHO2}. A campaign of adding software-simulated signals to the 
data in order to estimate the $h_0$ that corresponds to a 95\% average detection rate was carried out. 
We quantify sensitivity using the \emph{sensitivity depth} \cite{PhysRevD.91.064007,PhysRevD.98.084058}
\begin{equation}
    \mathcal{D} = \frac{\sqrt{S_{\textrm{n}}}}{h_0}\;
    \label{eq:depth}
\end{equation}
where $S_{\textrm{n}}$ represents the single-sided Power Spectral Density of the data (PSD), 
$\sqrt{S_{\textrm{n}}}$ is referred to as the Amplitude Spectral Density (ASD) and $h_{0}$ is
the previously defined CW amplitude. This figure of merit characterizes the sensitivity of the search 
to putative signals and accounts for the detector sensitivity as a function of frequency.
The actual single-sided PSD in Eq.~\eqref{eq:depth} depends on the analysis method  being used. \bsh sensitivity 
is dominated by the first stage using the weighted number count statistic meaning one should use the 
\emph{inverse squared averaged} PSD as shown in equations (42) to (44) of \cite{PhysRevD.77.022001}
\begin{equation}
    S_\textrm{n}(f) = \sqrt{\frac{N_{\alpha}}{\sum_{\alpha} \left[S_{\alpha}(f)\right]^{-2}}}\;,
\end{equation}
where $S_{\alpha}(f)$ represents the running-median noise floor estimation using 101 bins corresponding to the
SFT labeled by starting time $t_{\alpha}$ at frequency $f$. 
The goal is to characterize the average detection rate by numerically computing the efficiency distribution with 
respect to the depth. The result is interpolated to find the estimated sensitivity depth that corresponds
to 95\% detection efficiency. Using Eq.~\eqref{eq:depth} the sensitivity depth is converted to the sensitivity
amplitude. It is in this last step where the systematic error of the calibration is potentially relevant.

Systematic error in the amplitude of calibration of C01 data (final calibration version) is estimated to be 
lower than $7\%$ ($68\%$ confidence interval) for both detectors over all frequencies throughout O3a 
\cite{Sun_2020}.  
Relative deviations of ASDs computed using C00 data with respect to ASDs computed using C01 data (used as a 
proxy for an estimate of systematic error in C00 data calibration which otherwise does not exist for all time 
or frequencies) are below 7\% for all frequency bands except in the $[59, 61]\;\textrm{Hz}$ sub-band, where the 
relative deviation is $10\%$.
Assuming the proxy for C00 systematic error is complete, the impact of such 10\%-level of systematic error 
is negligible to the conclusions of this analysis.

Five representative frequency bands are selected across each $25~\textrm{Hz}$ band and binary 
parameter space region, and five sensitivity depth values used, namely $[18, 20, 22, 24, 26] \;\text{Hz}^{-1/2}$. 
Two hundred signals drawn from uniform distributions in phase and amplitude parameters are added to the data
at each depth, band and binary parameter space region. For each simulated signal, \bsh analyzes the data again 
in order to evaluate how many of them are detected. Sensitivity depth values are selected such that the $95\%$  
efficiency depth was properly bracketed; regions H4B and H5B required two extra depth values 
$[14, 16]\;\text{Hz}^{-1/2}$ to ensure this. Using a small number of frequency bands drastically reduces the 
computing cost of the sensitivity estimation procedure while yielding consistent results when compared to an
exhaustive injection campaign, as justified in \cite{BSHO2}.

Three criteria must be fulfilled in order to label a simulated signal as ``detected''. First, the toplist
obtained from the injection search should contain at least one candidate whose significance Eq.~\eqref{eq:significance}
is greater than the minimum significance present in the corresponding all-sky toplist. 
Second, after clustering the injection toplist, at least one cluster with a significance greater than the 
lowest significance recovered by the corresponding all-sky clustering must be obtained.
These two criteria ensure the injection is
prominent enough so as not to be discarded by the first stage of the search. Lastly, we require at least
one of the top five clusters from the injection toplist to be located closer than two parameter space bins in each
of the parameters with respect to the injection parameters. This last criterion takes into account the fact that,
in the actual search, a follow up will be done in corresponding regions around each significant cluster center.

\begin{figure}
    \includegraphics[width=0.5\textwidth]{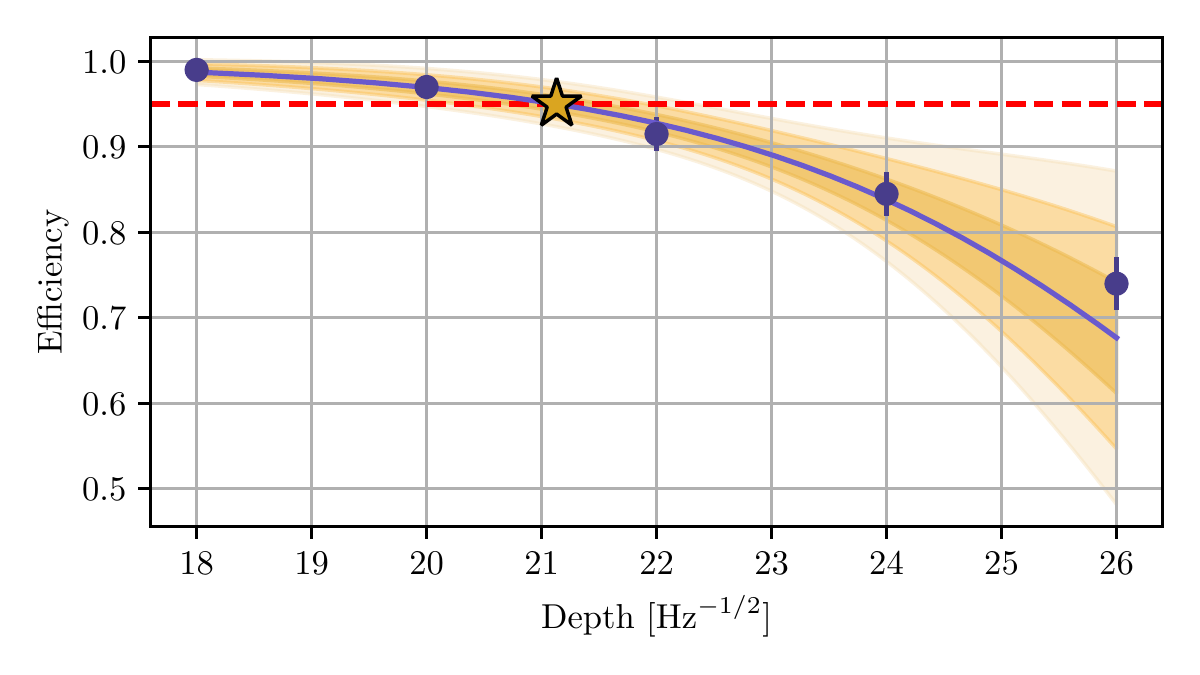}
    \caption{Example of 95\% sensitivity depth interpolation. Five sensitivity depths were selected at
    $124.625\;\textrm{Hz}$ in region H1B. 200 injections were injected at each of these depths, applying the
    criteria exposed in the text in order to label injections as detected/not detected. Blue dots represent the
    fraction of detected injections; the sigmoid fit is represented by a blue line; fit uncertainties at one, 
    two and three sigmas are represented by pale yellow shades. The interpolated $95\%$ sensitivity depth 
    $\mathcal{D}^{95\%} = 21 \pm 0.4$ is marked using a star.}
    \label{fig:interpolation}
\end{figure}

After separating detected from non-detected simulated signals, we construct efficiency curves akin to the
example shown in Fig. \ref{fig:interpolation}. Each point is the fraction of simulated signals detected (i.e. 
detection efficiency) as a function of the sensitivity depth. For each sensitivity depth set of 
$N_{\textrm{I}} = 200$ simulated signals, the uncertainty on detection efficiency $E$ is given by 
\begin{equation}
    \delta E = \sqrt{\frac{E \cdot \left(1 - E \right)}{N_\mathrm{I}}}\;.
\end{equation}

Then, using SciPy's \texttt{curve\_fit} function \cite{2020SciPy-NMeth}, we fit a sigmoid curve to the data
given by
\begin{equation}
    S(\mathcal{D}; \vec{p}\,) = 1 - \frac{1}{1 + e^{-p_{0}(\mathcal{D} - p_{1})}}\;,
\end{equation}
with fitted parameters $\vec{p}= \left(p_0, p_1\right)$. This expression can be inverted in order to find
the $95\%$ sensitivity depth.

\begin{figure}
    \includegraphics[width=0.49\textwidth]{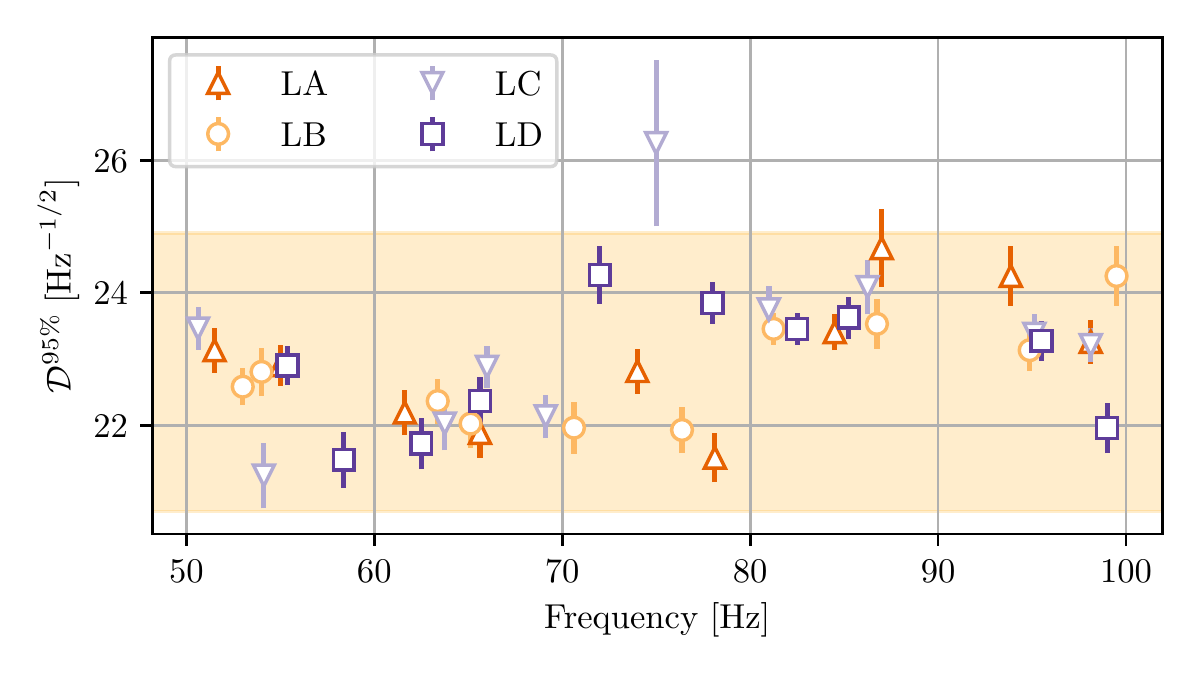}
    \includegraphics[width=0.49\textwidth]{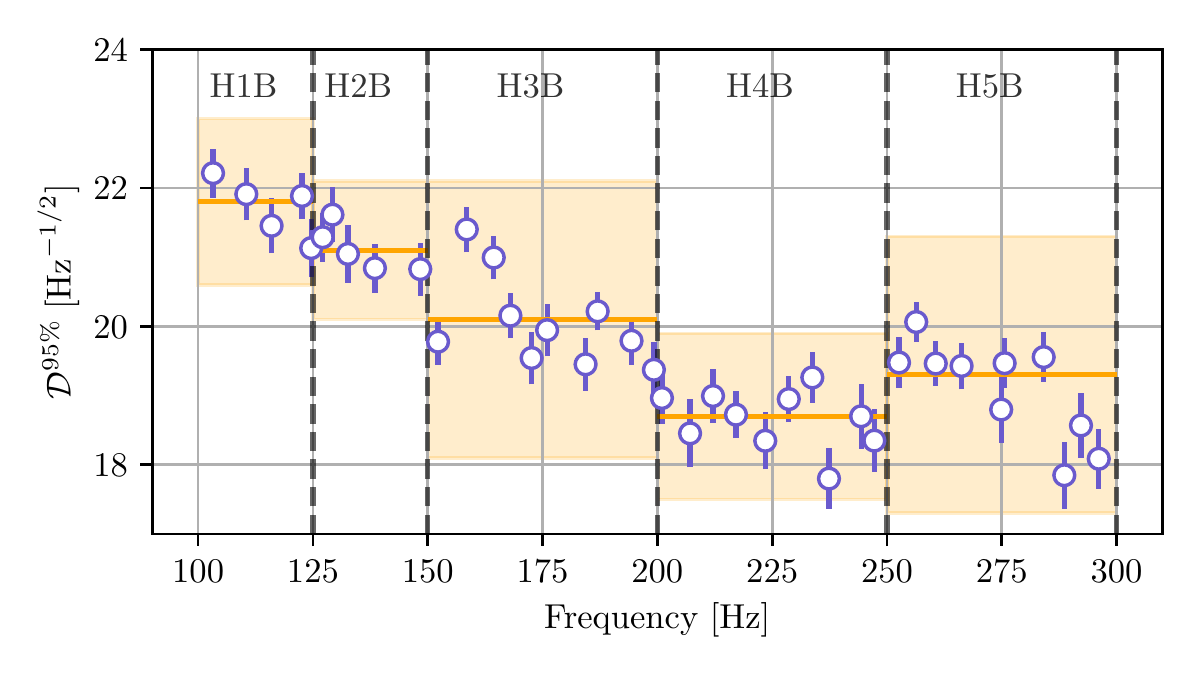}
    \caption{Average $95\%$ sensitivity depths obtained in the low-frequency (top panel) and high-frequency
    (bottom panel) bands. Data points correspond 
    to the interpolated results obtained through the sigmoid fit of the efficiencies at the selected
    frequency bands (5 bands randomly selected in each 25~Hz). Error bars correspond to 95\% efficiency
    uncertainties towards low depth values. Shaded regions show the averaged results with their uncertainties, 
    as summarized in Table~\ref{table:average_depths}.
    In the top panel, shading is only shown for the results obtained for the binary parameter space region B.
    }
    \label{fig:depth_plots}
\end{figure}

The interpolations are accompanied by a corresponding uncertainty, obtained through the covariance matrix
of the fit $\mathcal{C}(\mathcal{D})$ as
\begin{equation}
    \delta \mathcal{D}^{95\%} =
    \left.\sqrt{\vec{p}\,^{\mathrm{T}} \cdot 
    \mathcal{C}(\mathcal{D}) \cdot \vec{p}\,}
    \right|_{\mathcal{D}=\mathcal{D}^{95\%}}\;.
\end{equation}
The resulting interpolated depths per frequency band are shown in Fig. \ref{fig:depth_plots}. The high-frequency
search shows a clear degradation of depth values as frequency increases. This is related to 
the decaying density of parameter space templates: the higher the frequency, the finer one must construct a
template bank in order to achieve a comparable level of sensitivity. 

\begin{figure*}
    \includegraphics[width=\textwidth]{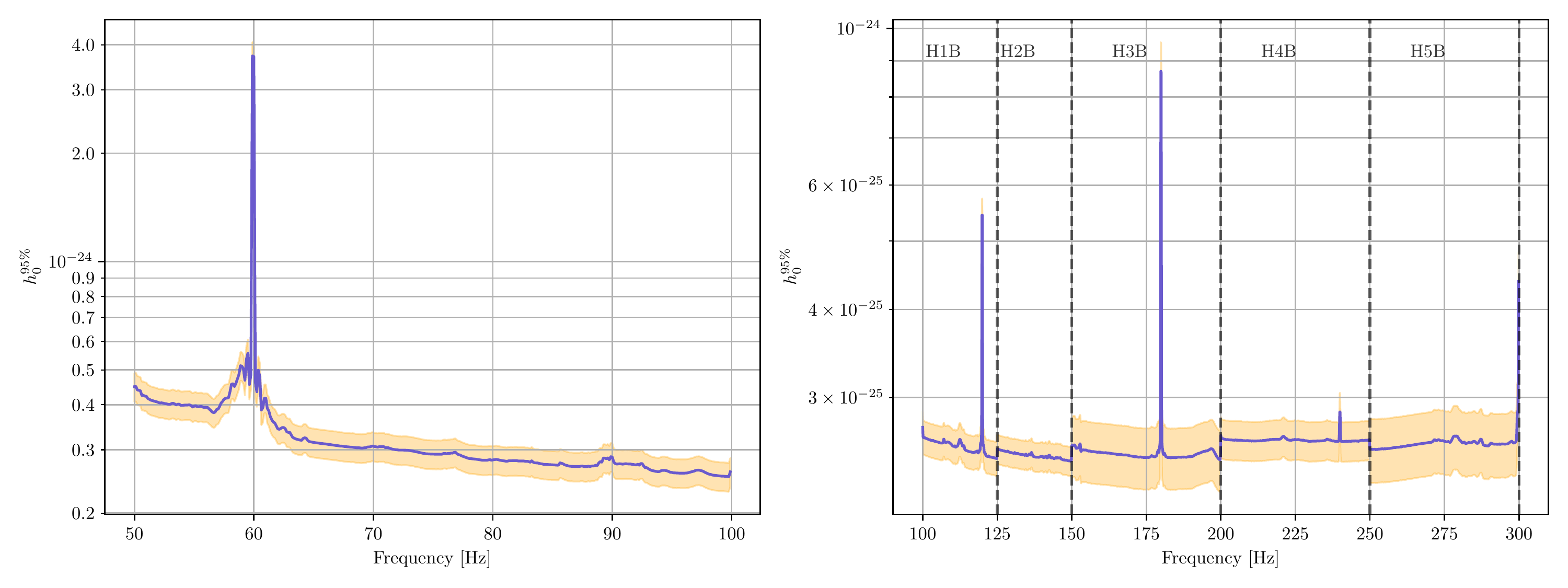}
    \caption{Implied $95\%$ efficiency amplitude from the obtained sensitivity depth values. The $h_{0}^{95\%}$ 
    amplitude estimates are obtained from the $95\%$ efficiency depth values shown in
    Table~\ref{table:average_depths} and the \emph{inverse squared averaged} PSD using
    Eq.~\eqref{eq:depth}. Low-frequency results are shown for binary parameter space region B.}
    \label{fig:sensitivity_amplitude}
\end{figure*}

Finally, we compute an average $95\%$ sensitivity depth for each of the regions quoted in 
Table~\ref{table:average_depths}. We also quote a corresponding $3\sigma$ uncertainty,
which previous studies have proven to deliver a good coverage of the actual $95\%$ efficiency sensitivity 
depth \cite{BSHO2}. These values are translated to CW amplitude $h_0$ via Eq.~\eqref{eq:depth} and shown 
in Fig. \ref{fig:sensitivity_amplitude}.


\begin{table}[]
\begin{tabular}{lc}
    \toprule
    Region  & $\langle \mathcal{D}^{95\%} \rangle \pm 3\sigma \;[\text{Hz}^{-1/2}]$\\ 
    \midrule
    LA & $22.9 \pm 2.5$ \\
    LB & $22.8 \pm 2.1$ \\
    LC & $23.0 \pm 2.5$ \\
    LD & $23.0 \pm 2.5$ \\
    H1B & $21.8 \pm 1.2$\\
    H2B & $21.1 \pm 1.0$\\
    H3B & $20.1 \pm 2.0$\\
    H4B & $18.7 \pm 1.2$\\
    H5B & $19.3 \pm 2.0$\\
    \bottomrule
\end{tabular}
    \caption{Average $95\%$ sensitivity depths for the parameter space regions analyzed in this search.
    Region labels are defined in Tables~\ref{table:binary_regions} and \ref{table:frequency_regions}.}
    \label{table:average_depths}
\end{table}

\section{Follow Up}\label{sec:fu}

Remaining candidates from the main search are followed up applying a more sensitive method to the data.
Longer coherence times constrain the phase evolution of the candidate under consideration and would yield 
higher significance for a true continuous wave signal. A potential downside remains that a true signal could 
be discarded if it is not well modeled by the assumed phase evolution. Moreover, increasing the coherence 
time also requires increasing the density of templates so as not to overlook a putative signal.

An effective way to cover small parameter regions is through Markov Chain Monte Carlo (MCMC) methods, 
which, rather than following a prescribed parameter space grid, sample the parameter space following
a certain probability density function. Reference \cite{Ashton:2018ure} describes how this can be 
implemented in a search for continuous waves by using the so-called $\mathcal{F}$-statistic, a 
well-established CW analysis technique, as a likelihood function. We refer to \cite{Ashton:2018ure}
and references therein for an in-depth explanation of this method.

The $\mathcal{F}$-statistic is a coherent statistic, usually referred to as $2 \tilde{\mathcal{F}}$,
which compares data against templates by matched filtering. A semicoherent $\mathcal{F}$-statistic 
$2 \hat{\mathcal{F}}$ can be defined by adding individual $2\tilde{\mathcal{F}}$ values computed over 
$N_{\textrm{seg}}$ segments spanning $T_{\textrm{coh}}$ each, in the same way as weighted normalized power
was computed from weighted power in Eq. \eqref{eq:wnormpower}:
\begin{equation}
    2 \hat{\mathcal{F}}(\lambda) = \sum_{s = 0}^{N_{\textrm{seg}}-1} 2 \tilde{\mathcal{F}}_{s}(\lambda)\;,
\end{equation}
where the index $s$ indicates the coherent quantity has been computed for a certain segment spanning
$T_{\textrm{coh}}$.

We use software injections in order to calibrate a threshold $2\hat{\mathcal{F}}_{\textrm{th}}$. 
Candidates such that $2\hat{\mathcal{F}}(\lambda) < 2\hat{\mathcal{F}}_{\textrm{th}}$ will be deemed as 
non-significant and consequently discarded.

This algorithm is implemented in \texttt{PyFstat} \cite{pyfstat, pyfstat_zenodo}. It builds on top of LALSuite 
\cite{lalsuite}, which provides the CW data analysis functionality, and \texttt{ptemcee} 
\cite{Foreman_Mackey_2013, 10.1093/mnras/stv2422}, which implements the MCMC algorithms.

\subsection{MCMC follow-up configuration}\label{sec:priors}

\begin{table}[]
\begin{tabular}{ll}
    \toprule
    Hyperparameter & Value  \\ 
    \midrule
    Parallel chains & 3              \\
    Walkers per chain & 100            \\
    Burn-in \& Production steps & 100 + 100  \\ 
    \bottomrule
\end{tabular}
    \caption{MCMC hyperparameter choice for the first stage of the follow up. Each of the 
    \emph{parallel chains} sample the likelihood at a different temperature, as explained in
    \cite{10.1093/mnras/stv2422}.}
    \label{table:hyperparameters}
\end{table}
The MCMC follow up employed is not intended to describe the posterior distribution of parameters 
defining a candidate. Rather, we only require enough convergence such that the sampled $\mathcal{F}$-statistic 
values are close enough to the local maximum to establish a reliable veto threshold.

\subsubsection{Sampler configuration}

The \texttt{ptemcee} package implements an ensemble-based sampler that uses several walker chains to sample 
multimodal distributions.
Expensive setups are not required in order to perform a first-stage follow up using a threshold-based approach. 
The reason for this is two-fold: we are increasing the coherence time with respect to the search, and we do not 
require extensive convergence to be achieved. No second-stage follow up was required because all of the 
first-stage outliers were attributed to instrumental causes. If this was not the case, we would have applied 
a second follow-up stage using a more expensive setup. The number of parallel chains, walkers per chain, 
and number of steps to take are summarized in Table~\ref{table:hyperparameters}.

We choose to use $N_{\textrm{seg}} = 260$, which corresponds to $T_{\textrm{coh}} \simeq 17\,\textrm{h}$.
This is a longer coherence time with respect to that of the initial stage of the search, a choice used in 
previous searches \cite{BSHO2}.

\subsubsection{Prior choice}

Following a similar prescription as the one given in \cite{Ashton:2018ure}, we set up uniform
priors in each parameter space dimension, forming a box centered on each cluster center. Each edge of
this box spans two parameter space bins according to the spacing given in Eq.~\eqref{eq:resolution}, 
where the parameter-space-dependent quantities are computed at the center of the cluster. This is in 
agreement with the detection criteria imposed to perform the sensitivity estimation.

Although \bsh targets CW sources in circular orbits, it is still sensitive
to signals with eccentricities up to a certain value, as long as the Doppler modulation derived from eccentricity 
is smaller than half a frequency bin. The upper bound for the maximum allowable eccentricity according to 
this argument was derived in \cite{BSH}
\begin{equation}
    e^{\textrm{m.a.}} = \left[2 \, \Tsft \, f_{0} \, a_p\,\Omega \right]^{-1} \;.
\end{equation}
Therefore, uniform priors on eccentricity, $[0, e^{\textrm{m.a.}}]$, and argument of periastron, $[0, 2\pi]$ 
are included as MCMC parameters.
Maximum eccentricities range from $0.2-0.5$ at $50~\textrm{Hz}$ to less than $0.1$ at $300~\textrm{Hz}$.

\subsection{Setting up a threshold}

\begin{figure}
    \includegraphics[width=0.49\textwidth]{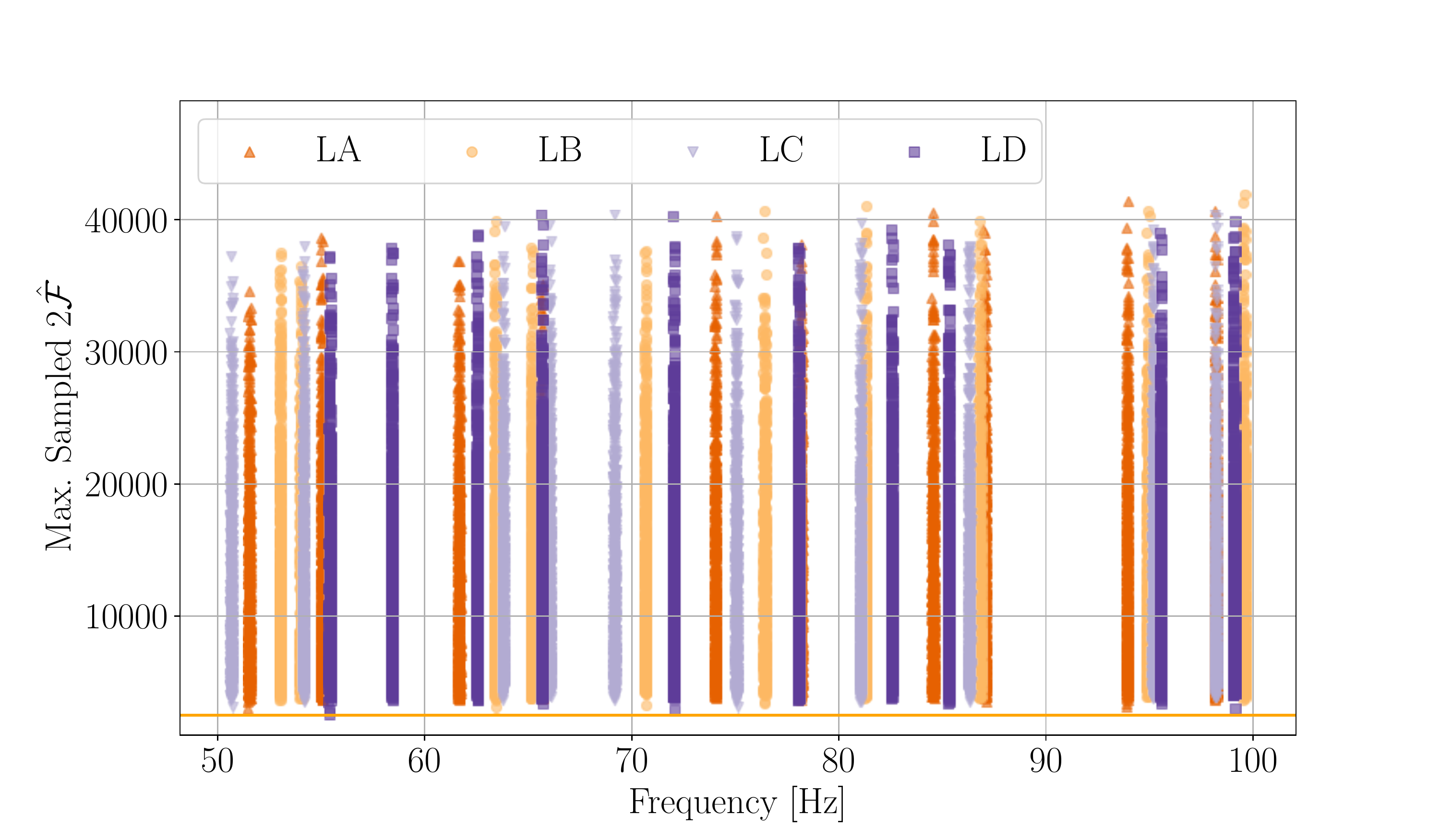}
    \includegraphics[width=0.49\textwidth]{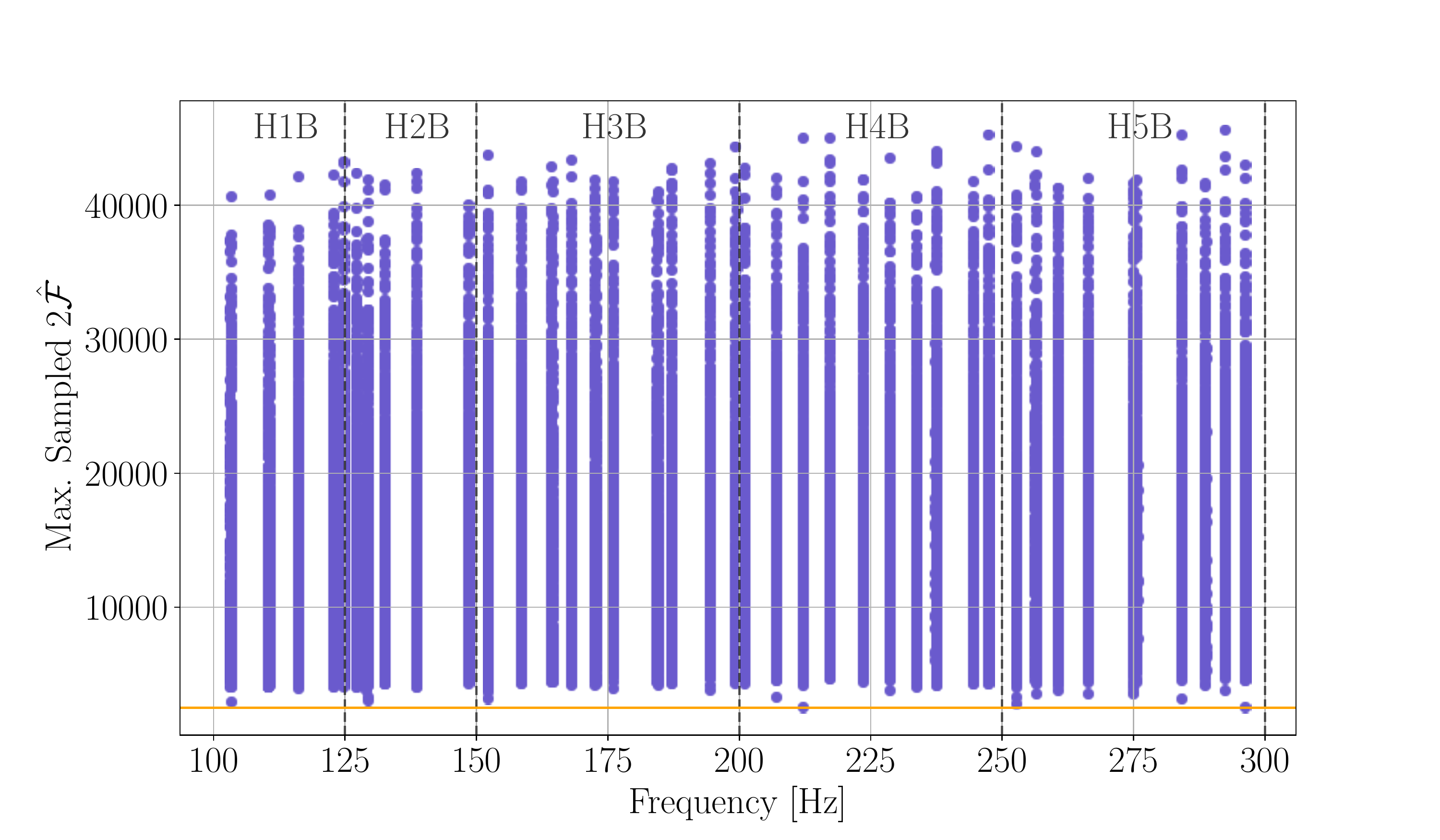}
    \caption{Recovered $\fsemi$ values using a set of injections labeled as \emph{detected} by the 
    sensitivity estimation criteria. The amplitudes of these injections were distributed using the five 
    sensitivity depth values explained in Sec.~\ref{sec:ul}.
    The top panel shows the results for the four regions of the
    low-frequency search, involving 33757 injections; the bottom panel shows the same result for the 
    high-frequency search, using 37549 injections. The horizontal orange line marks the threshold 
    $2\hat{\mathcal{F}}_{\textrm{th}} = 2500$.}
    \label{fig:follow_up_threshold}
\end{figure}

We use the \bsh and the MCMC follow up on a total of 71306 software injections in order to calibrate 
a significance threshold. The employed injections are consistent with the ones used for the sensitivity 
estimation, focusing on those detected according to the three criteria (see Sec. VI). This implies a 
significant fraction of the injections will possess an amplitude below the obtained $95\%$ sensitivity 
amplitude, as they will be distributed according to the five original depths. The threshold obtained using
this calibration strategy will have a low false-dismissal rate 
\mbox{($\lesssim 1/71306 \simeq 1.5 \times 10^{-5}$)} against signals detectable by this pipeline.

We run the MCMC algorithm in order to sample $2\hat{\mathcal{F}}$ values, retrieving the maximum value for
each of the injections. Resulting $\fsemi$ values for these simulations are plotted in Fig. 
\ref{fig:follow_up_threshold}. These results support the choice of $2\hat{\mathcal{F}}_{\textrm{th}} = 2500$ 
as the threshold value, with all detected injections above this threshold.

\subsection{Surviving Outliers}\label{sec:surviving}

\begin{figure}
    \includegraphics[width=0.49\textwidth]{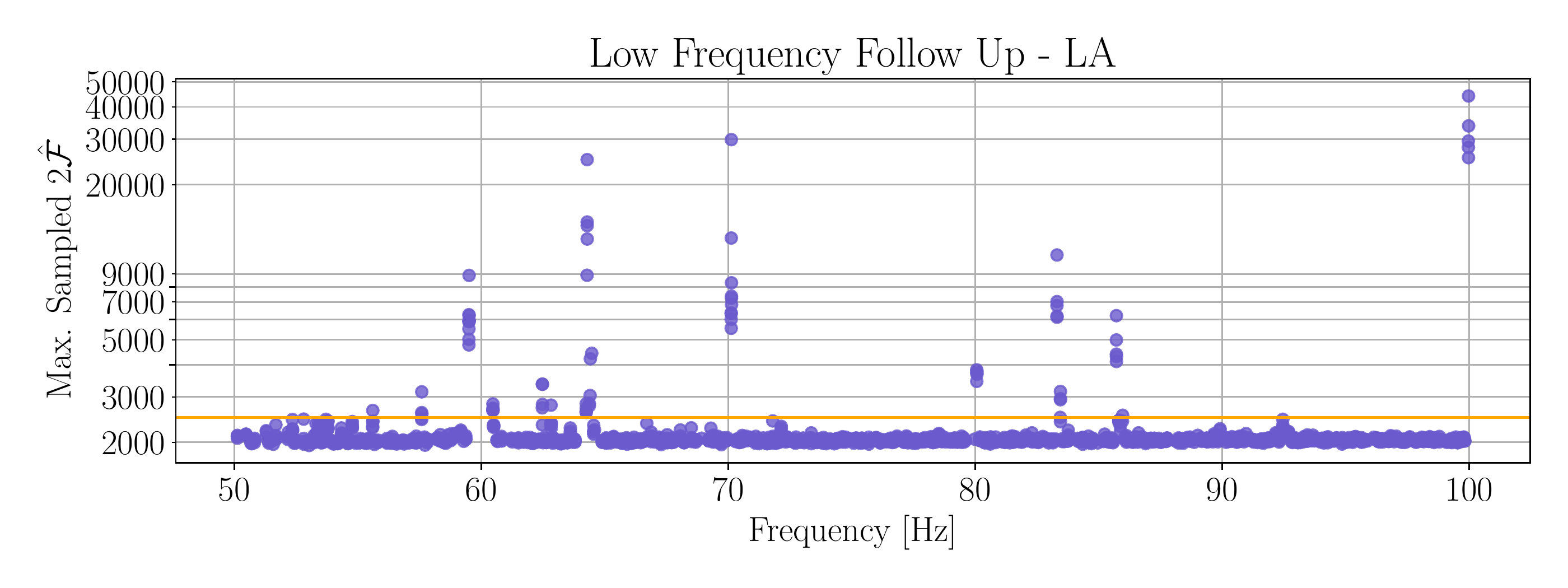}
    \includegraphics[width=0.49\textwidth]{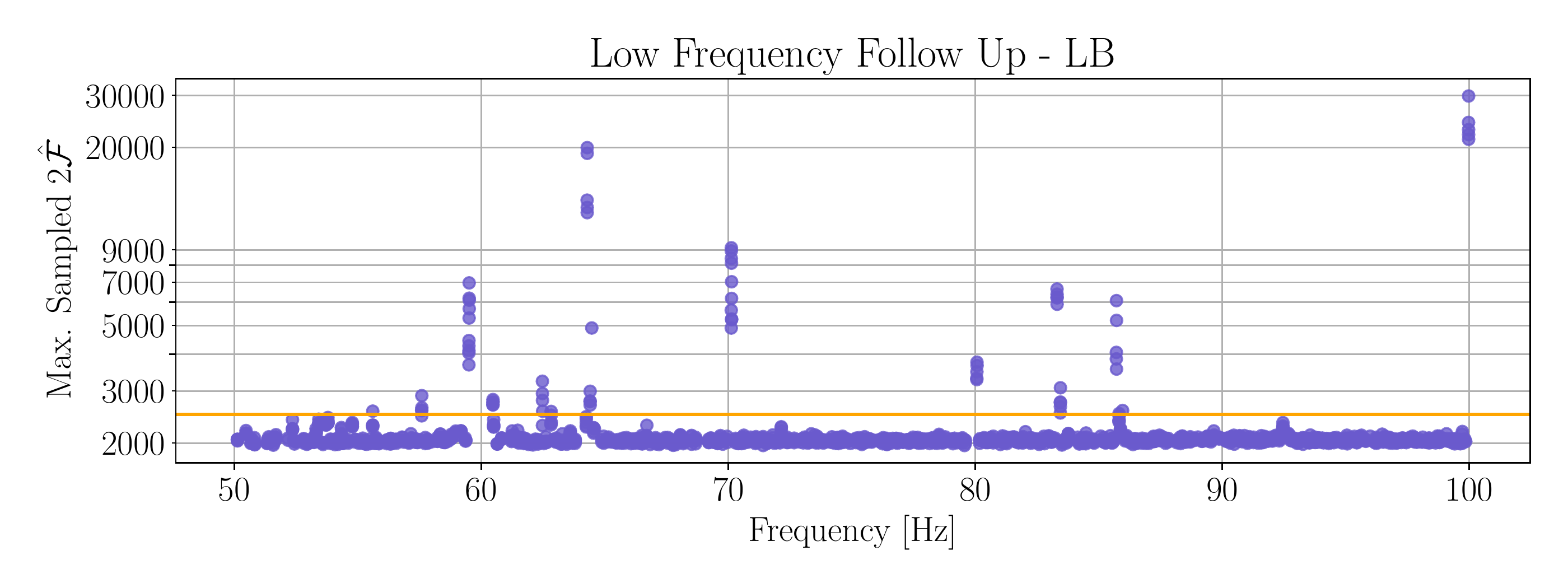}
    \includegraphics[width=0.49\textwidth]{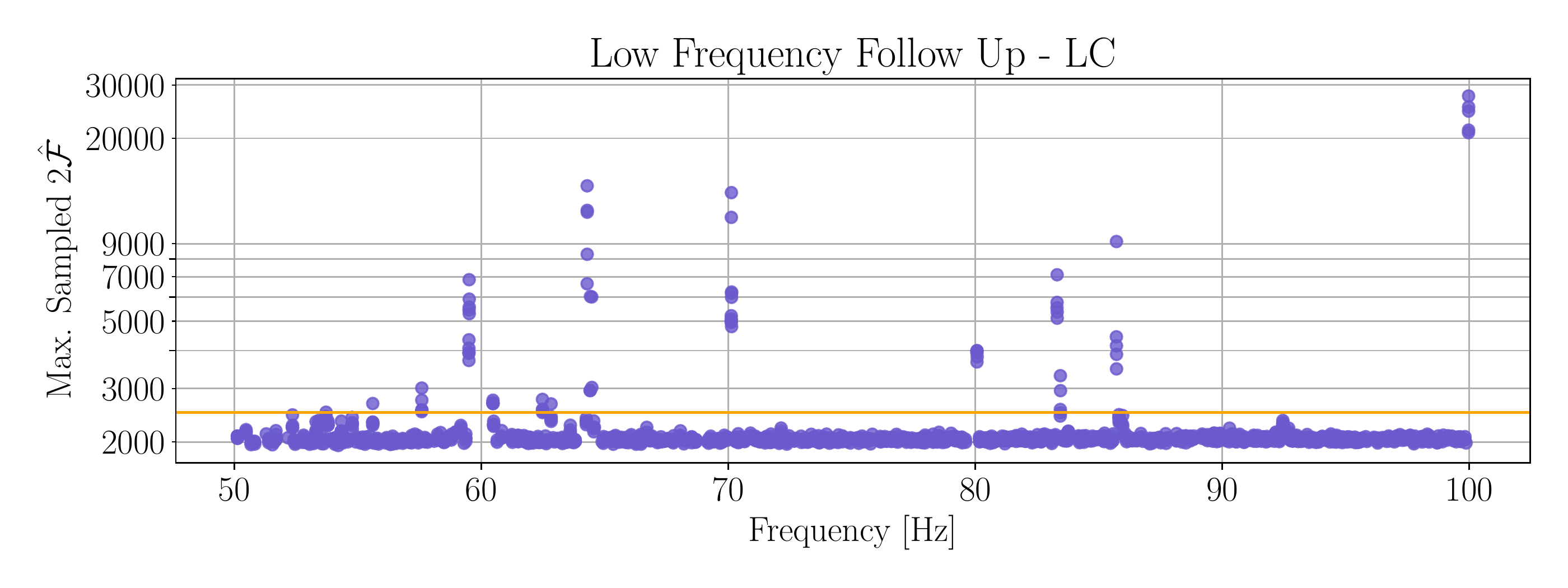}
    \includegraphics[width=0.49\textwidth]{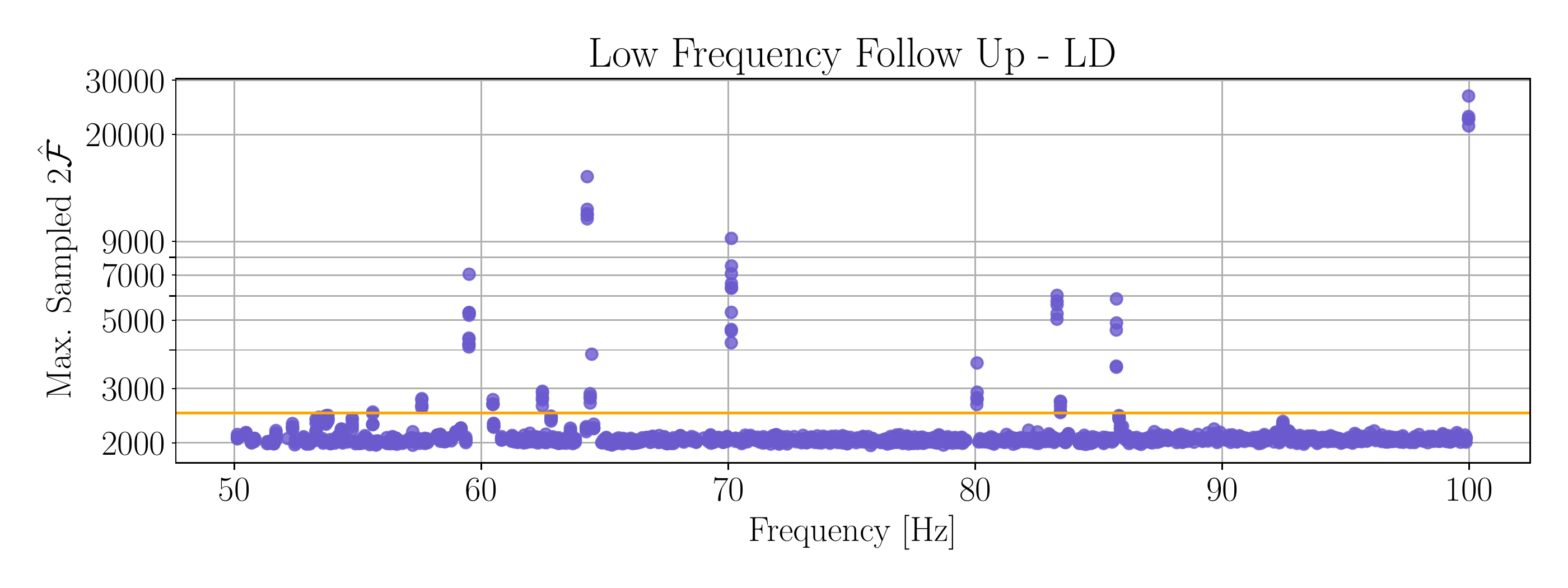}
    \includegraphics[width=0.49\textwidth]{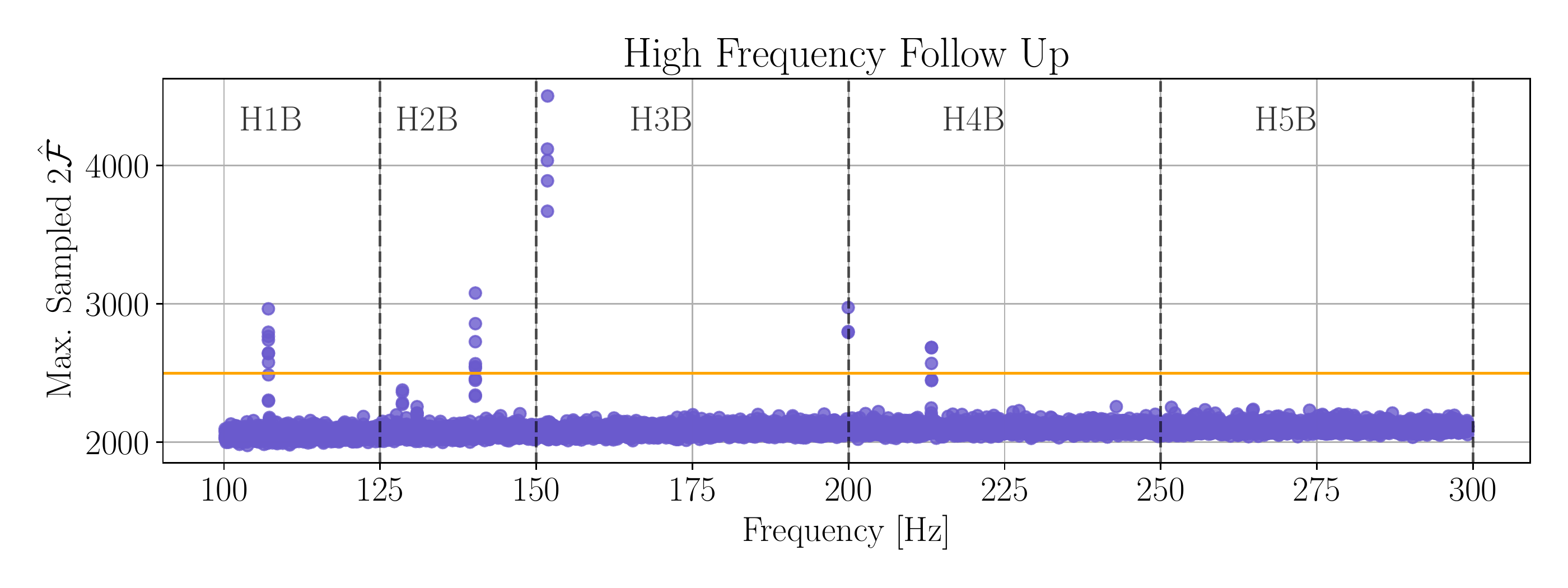}
    \caption{MCMC follow-up results. Dots represent cluster centers not vetoed
    by the identified line veto, and a horizontal line represents the imposed threshold
    $\fsemi_{\textrm{th}}=2500$,
    calibrated by software injections (see Fig. \ref{fig:follow_up_threshold}). The horizontal axis represents 
    the frequency value associated to each cluster center, and the vertical axis represents the maximum $\fsemi$
    value sampled by the MCMC run.}
    \label{fig:follow_up_results}
\end{figure}

After executing the MCMC follow up and imposing the $2\hat{\mathcal{F}}_{\textrm{th}} = 2500$ threshold, 
287 outliers remain in the low-frequency band and 24 outliers remain in the high-frequency band, as shown in 
Fig. \ref{fig:follow_up_results}. It is clear from the figure that low-frequency outliers mostly belong
to the same frequency bands across the four binary parameter space regions.  
We next analyze each candidate using a cumulative semicoherent $\mathcal{F}$-statistic, defined as
\begin{equation}
    \fsemi(\lambda; t) = \sum_{\alpha : t_{\alpha} < t} 2 \tilde{\mathcal{F}}_{\alpha}(\lambda)\;,
    \label{eq:cumF}
\end{equation}
in order to discern those candidates originating from instrumental noise.

We use three flavors of Eq.~\eqref{eq:cumF}, one for each of the detectors (H1 and L1)
and another one using a multi-detector approach (H1 + L1). These statistics lead to the rejection of the 
remaining outliers, as described below.

\subsubsection{Line-crossing outliers}
\label{sec:transient_lines}

\begin{figure}
    \includegraphics[width=0.5\textwidth]{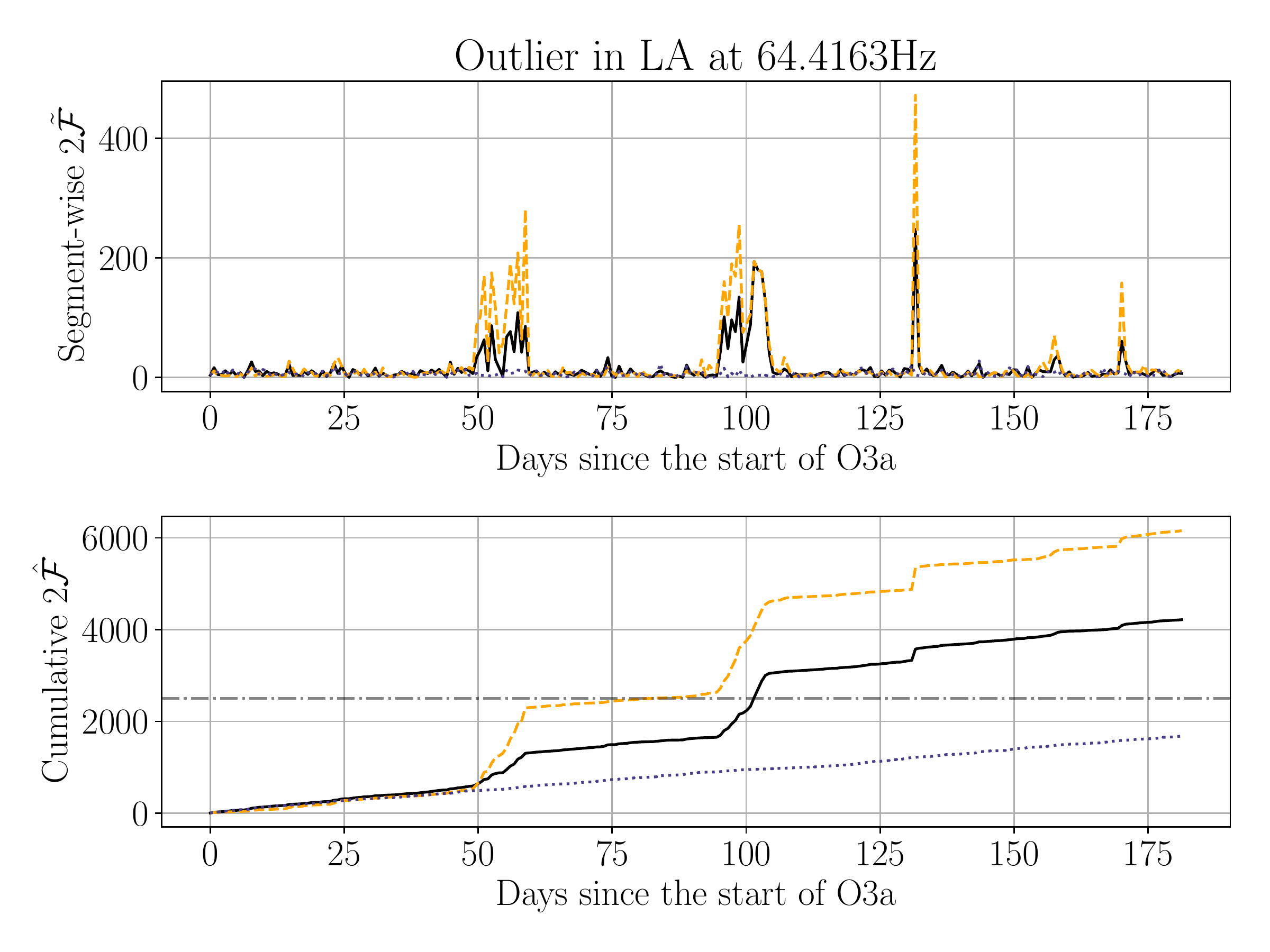}
    \includegraphics[width=0.5\textwidth]{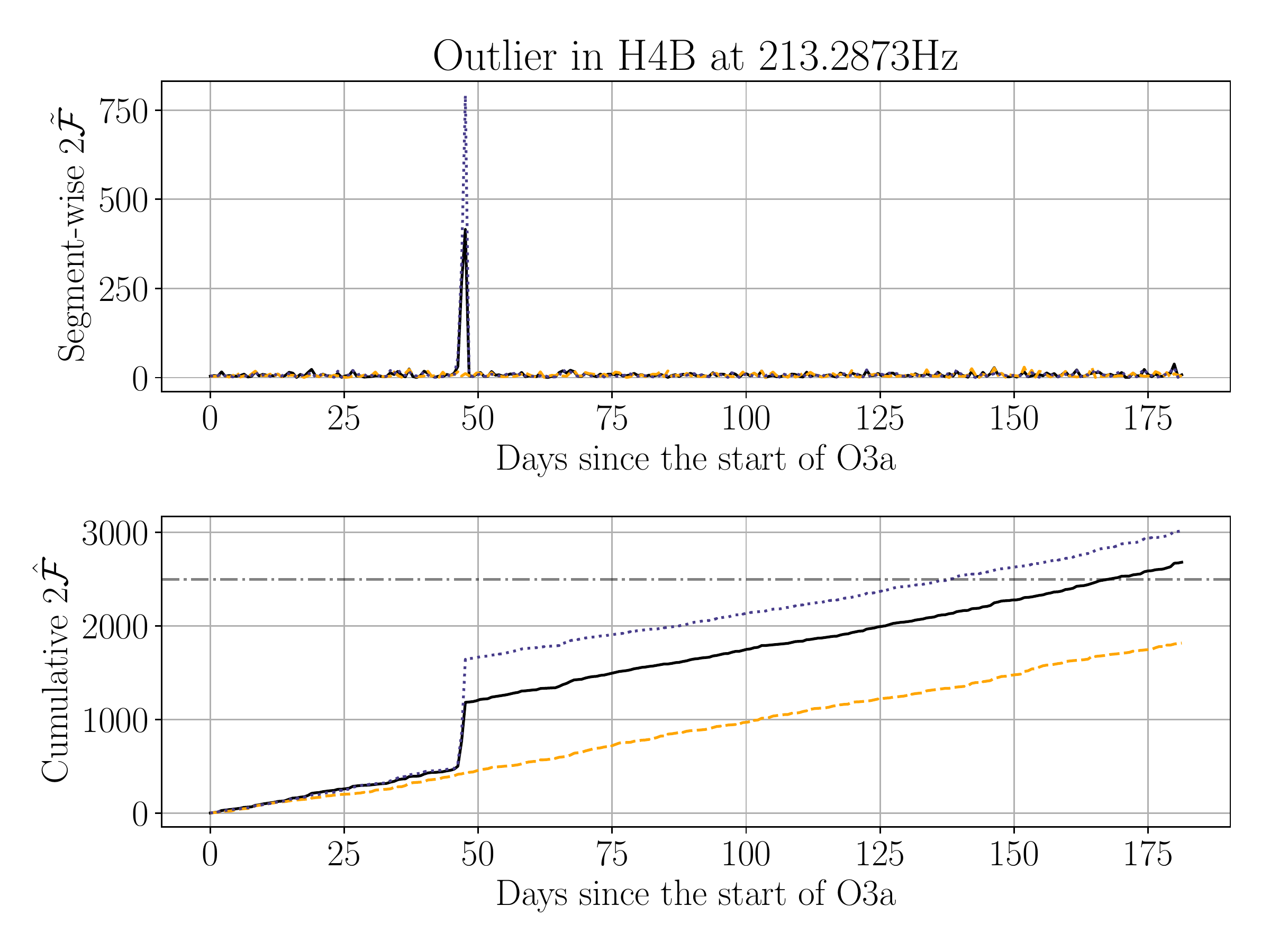}
    \caption{Example of outliers produced by instrumental noise in one of the Advanced LIGO detectors.
    Each pair of panels corresponds to a different outlier; the segment-wise $2 \tilde{\mathcal{F}}_{\alpha}$ 
    statistic is shown at the top of each pair; the cumulative semicoherent $\fsemi$ as described in 
    Eq.~\eqref{eq:cumF} is shown at the bottom of each pair. Dashed lines denote an H1-only analysis, 
    dotted blue lines an L1-only analysis, and solid black lines a multi-detector analysis.
    The $\fsemi_{\textrm{th}}=2500$ threshold is shown as a horizontal line.}
    \label{fig:glitch}
\end{figure}

CWs are expected to accumulate a $\fsemi$ value linearly with respect to the observing time.
We find that 263 outliers surpass the $\fsemi_{\textrm{th}}=2500$ threshold due to the presence of 
prominent values of segment-wise $\mathcal{F}$-statistic at certain times of the run in one of the detectors 
(260 in H1 and 3 in L1), as exemplified in Fig. \ref{fig:glitch}. The higher number of outliers in H1
arise from the greater number of instrumental lines present in that detector \cite{IdentifiedLines, 
UnidentifiedLines}.

This behavior would be expected from a candidate whose frequency evolution track crosses a narrow instrumental 
artifact (line) for a limited duration, either because of the frequency track drifting away from the line
or the transient nature of the line itself. Most strong persistent instrumental disturbances are already 
discarded using the known lines list \cite{IdentifiedLines}, but weaker lines or transient disturbances 
(lasting hours to days), which are more difficult to identify in the run-averaged spectra, could still affect 
our searches \cite{Keitel:2015ova, PhysRevD.97.082002}.

In Table~\ref{ref:spiky_outliers} in Appendix \ref{appendix}, we present a list of frequency bands containing 
these 263 candidates whose behavior suggests a brief line crossing or the presence of a transient instrumental 
disturbance. Outliers were selected as belonging to this category if they have at least one per-segment 
$\mathcal{F}$-statistic value greater than $100$, and for each frequency band listed in the table, the 
first/last timestamps bracket the data segments where $\mathcal{F}$-statistic values greater than $50$ were 
observed for at least one of those candidates. Overlapping frequency bands were merged together for the sake
of clarity.

\subsubsection{Detector consistency veto}
\label{sec:consistency_veto}

\begin{figure}
    \includegraphics[width=0.5\textwidth]{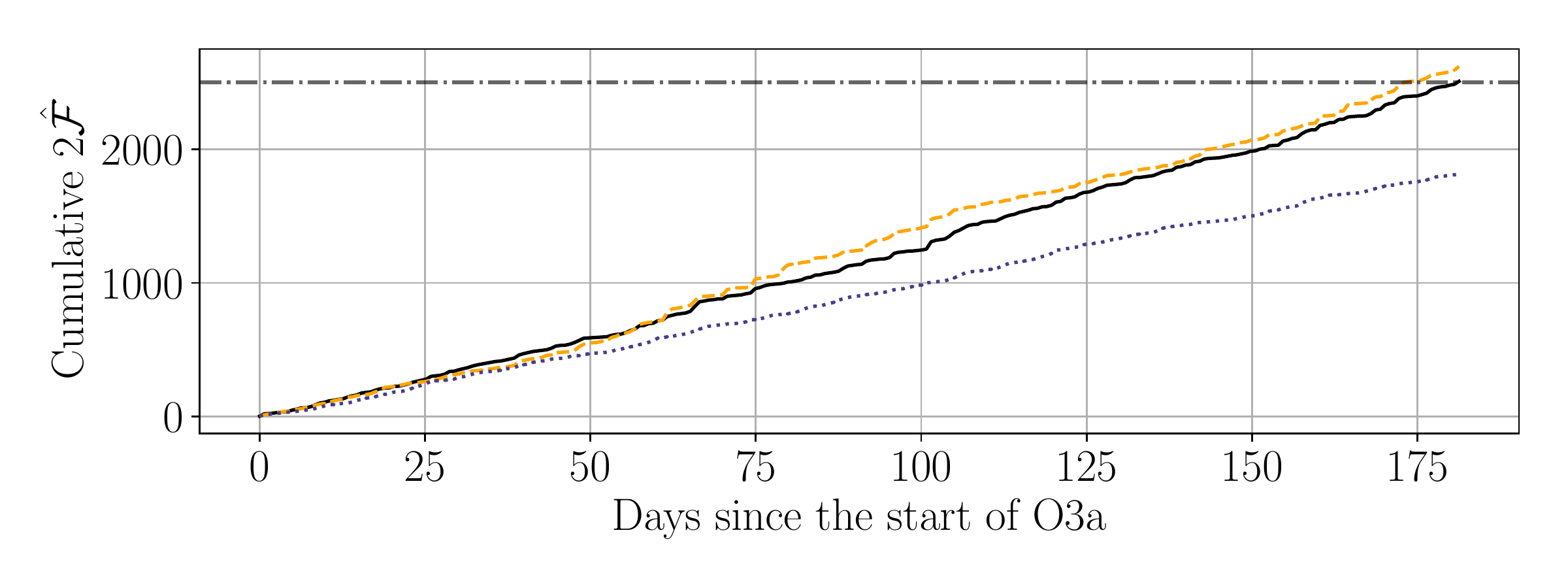}
    \caption{Example of an outlier vetoed by the multi-detector consistency veto. 
    Dashed lines denote an H1-only analysis, 
    dotted blue lines an L1-only analysis, and solid black lines a multi-detector analysis.
    The $\fsemi_{\textrm{th}}=2500$ threshold is shown as a horizontal line.}
    \label{fig:instrumental_noise}
\end{figure}
\begin{figure}
    \includegraphics[width=0.5\textwidth]{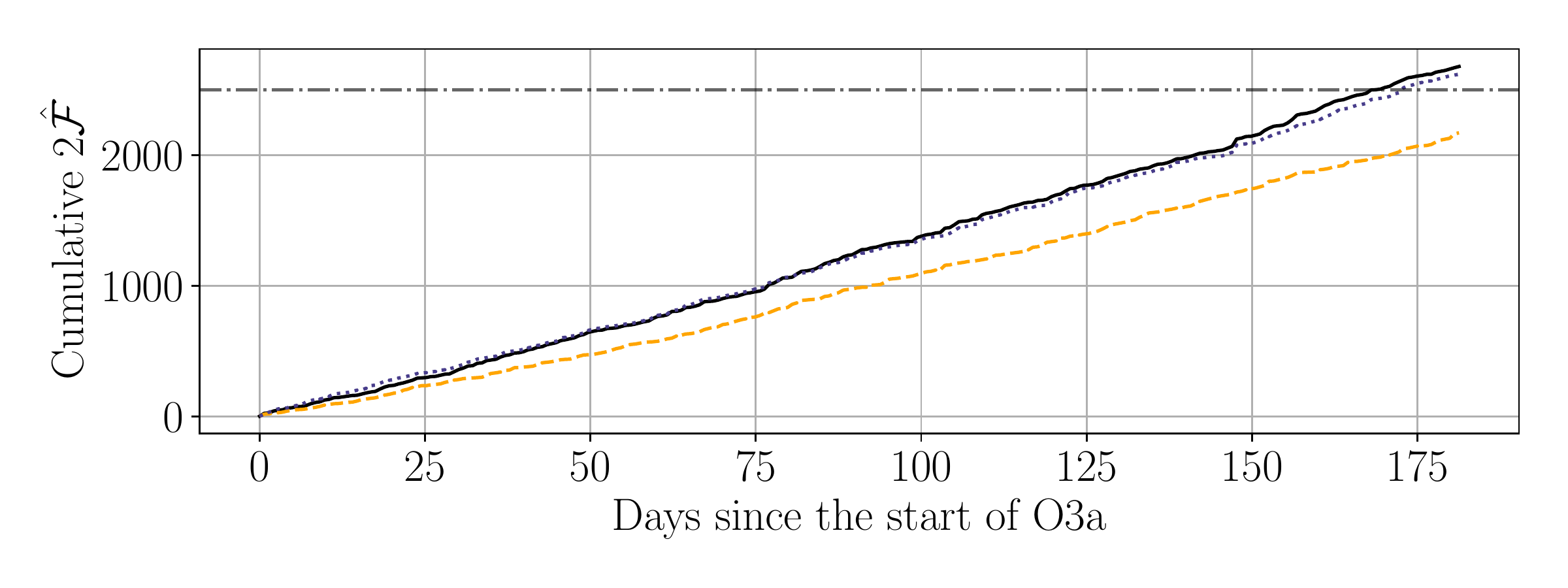}
    \caption{Example of an outlier surpassing the $\fsemi_{\textrm{th}}=2500$ threshold
    later vetoed by inspection of the detector spectra.
    The combined ASD of both detectors around the frequency of this group of outliers is shown in Fig.~\ref{fig:outlier_location}.
    Candidates related to this signal saturated the toplist in the four regions LA, LB, LC, LD.
    Dashed lines denote an H1-only analysis, 
    dotted blue lines an L1-only analysis, and solid black lines a multi-detector analysis.
    The $\fsemi_{\textrm{th}}=2500$ threshold is shown as a horizontal line.}
    \label{fig:interesting_candidates}
\end{figure}

A second set of 28 outliers is discarded by the detector consistency veto (see e.g. \cite{Abbott:2017mnu}). 
During O3a, the L1 detector presents a better sensitivity than H1 at low frequencies \cite{O3aRun}.
A CW candidate would be expected to behave consistently, i.e.
\mbox{$\fsemi_{\textrm{L1}} > \fsemi_{\textrm{H1}}$} for most signals. 
We calibrate this veto using the aforementioned set of software injections in order to take detector
sensitivity anisotropies due to the antenna pattern functions into account, obtaining a maximum relative $5\%$ 
excess of $\fsemi_{\textrm{H1}}$ with respect to $\fsemi_{\textrm{L1}}$.

The 28 outliers rejected with this veto show more than a 30\% relative excess of $\fsemi_{\textrm{H1}}$ with 
respect to $\fsemi_{\textrm{L1}}$. Hence, we discard them as being inconsistent with an astrophysical signal. 
Figure \ref{fig:instrumental_noise} shows an example of these outliers.
After computing the bandwidth covered by each of these candidates, we obtain seven distinct frequency 
bands affected by instrumental disturbances of this type, summarized in Table~\ref{ref:persistent_outliers} in
appendix \ref{appendix}. 

\subsubsection{Powerline sidebands}

\begin{figure}
    \includegraphics[width=0.5\textwidth]{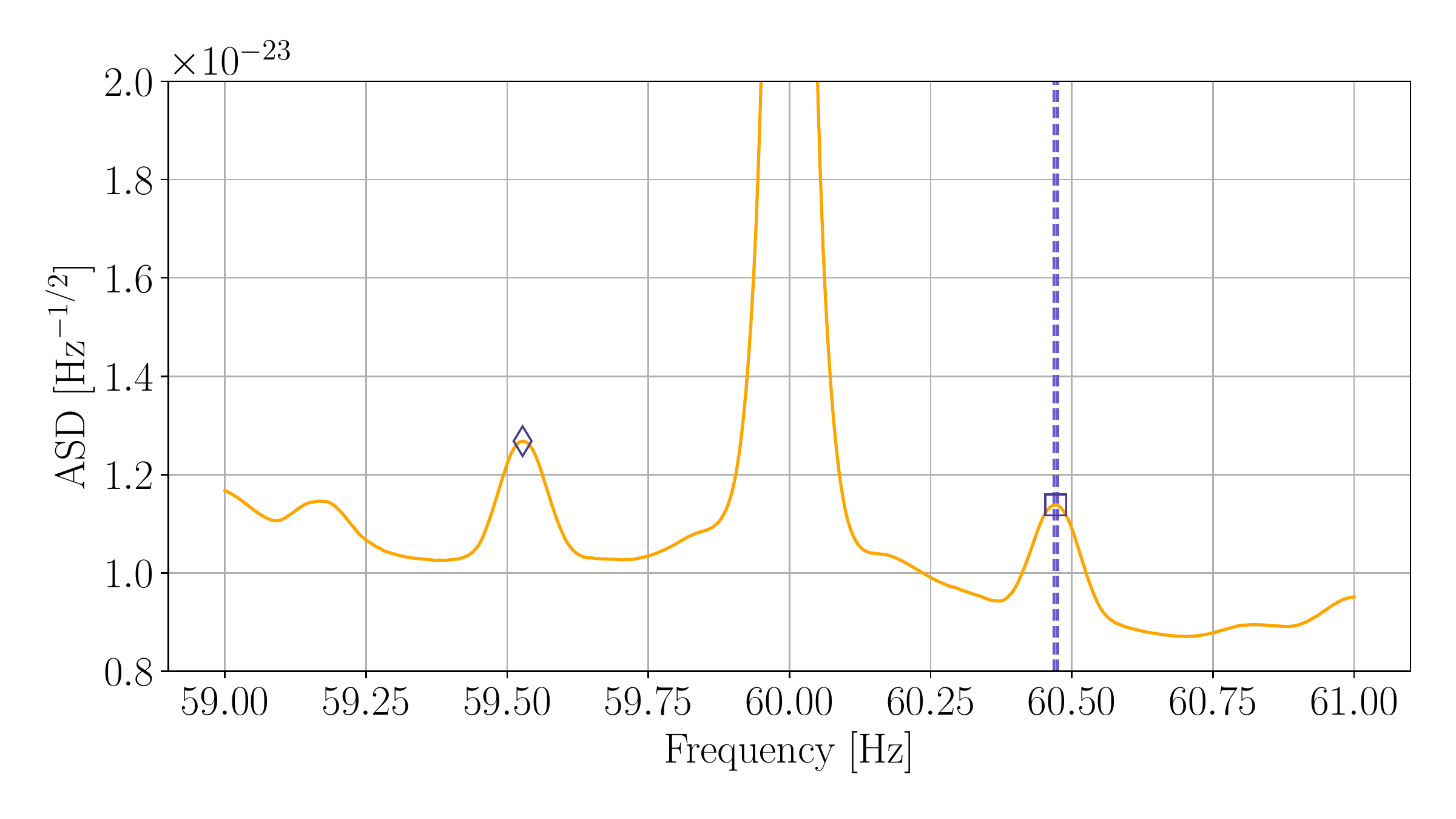}
    \caption{Location of the manually inspected outliers (vertical dashed line) with respect to the amplitude 
    spectral density of the detector, here represented as a multi-detector \emph{inverse squared average} 
    (orange line). A diamond and a square mark the twin peaks' frequency, $59.53\textrm{Hz}$ and 
    $60.47\textrm{Hz}$ respectively.
    }
    \label{fig:outlier_location}
\end{figure}

The last 20 outliers were consistently present in each one of the four parameter space regions within the 
$[60.46, 60.48]\;\textrm{Hz}$ sub-band. These outliers were not vetoed by any of the previous stages.
As shown in Fig. \ref{fig:interesting_candidates}, $\fsemi$ was accumulated in a fairly linear fashion,
achieving greater values in L1 than H1. The detector ASD (Fig. \ref{fig:outlier_location}), 
however, shows that these candidates were caused by sidebands of the $60~\textrm{Hz}$ power supply artifact.
These sidebands can be explained by a \mbox{non-linear} coupling between the main power supply frequency and a 
low-frequency noise. They do not appear in the line lists \cite{IdentifiedLines, UnidentifiedLines} as they do
not correspond to narrow spectral artifacts and their effect on CW searches is highly dependent on the search 
method. Due to the presence of said artifact in the data and the wide spread of the candidates obtained 
by our search across these bands, we deem this final set of candidates as non-astrophysical.

\section{Conclusion}\label{sec:con}

\begin{figure*}
    \includegraphics[width=0.49\textwidth]{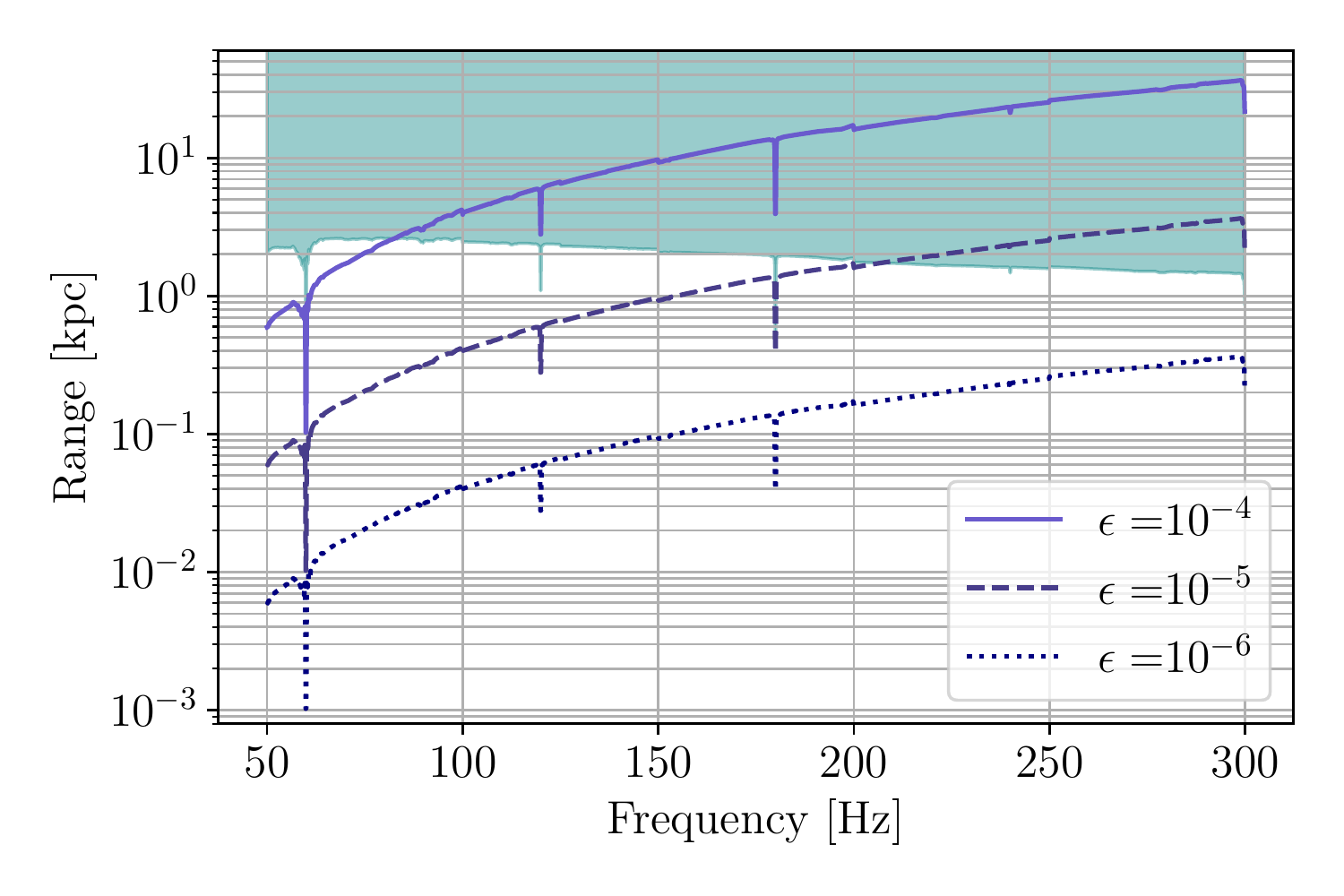}
    \includegraphics[width=0.49\textwidth]{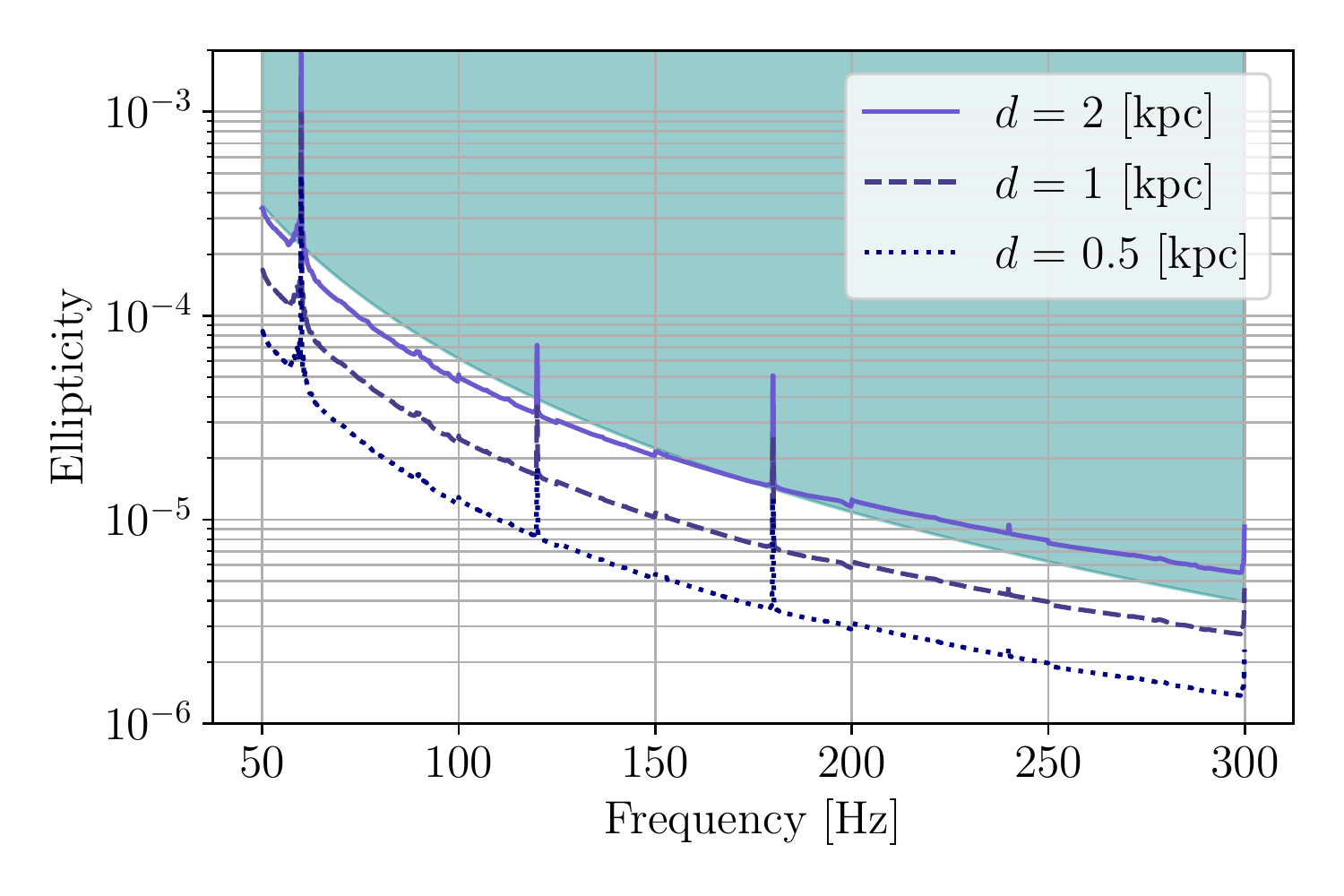}
    \caption{Astrophysical reach (left) and equatorial ellipticity (right) implied by the search sensitivity 
    $h_{0}^{95\%}$. These results were obtained assuming a canonical neutron star moment of inertia 
    \mbox{$I_{\textrm{z}} = 10^{38} \text{kg}\cdot \text{m}^2$}. Shaded areas denote regions excluded due to 
    the spindown limit implied by the maximum spindown value considered in this search.}
    \label{fig:reach}
\end{figure*}
We report on a search for continuous gravitational wave signals from unknown sources in binary systems using 
LIGO data from the first six months of the third Advanced LIGO and Advanced Virgo observing run. 
Four different binary parameter space regions, spanning orbital periods of $3 - 45\;\text{days}$ and 
projected semimajor axes of $2 - 40$ light-seconds, are searched across the $50 - 300\;\textrm{Hz}$ 
frequency band. We claim no detections and estimate the sensitivity of the search in terms of the 
gravitational wave amplitude corresponding to the interpolated $95\%$ detection efficiency using a 
simulated population of signals.

The minimum amplitude sensitivity attains an average value of $h_0^{95\%}= (2.4 \pm 0.1) \times 10^{-25}$ in 
the $f_{0} = 149.5\;\textrm{Hz}$ sub-band. This is a factor of $\sim\!1.6$ lower than the lowest amplitude
sensitivity obtained by a previous search performed on data from the second Advanced LIGO observing run 
\cite{BSHO2}. The estimated amplitude sensitivity can be interpreted in terms of astrophysical reach and 
equatorial ellipticity by means of equation Eq.~\eqref{eq:h0}, as shown in Fig. \ref{fig:reach}.

The validity of this estimation must be discussed in terms of the \emph{spindown limit}, which corresponds
to the maximum gravitational wave amplitude achievable by a neutron star assuming its rotational energy is 
solely lost via gravitational waves. We refer the reader to Appendix A of \cite{Abbott_2019} for its 
definition and the relevant conversion equations.

The maximum spindown value probed by our search is 
\mbox{$|\dot{f}_{0}| \equiv \left(\Tsft \cdot T_{\textrm{obs}}\right)^{-1} \simeq 6.5 \times 10^{-11} \textrm{Hz}/\textrm{s}$} \cite{BSH}, meaning sources braking at higher
rates would not be detected by our pipeline (see Table \ref{table:misc} for the definition of $\Tsft$ and 
$T_{\textrm{obs}}$). Assuming the canonical emission model of a deformed NS as in
Eq.~\eqref{eq:h0}, this implies the existence of a distance beyond which the required ellipticity to emit 
a detectable amplitude would imply a greater spindown than the one probed  by the search, as long as no 
processes balancing the rotational energy loss are in place\footnote{An accretion-driven torque balance 
\cite{Bildsten:1998ey} could be subject to fluctuating accretion \cite{Watts:2008qw}, leading to long-term 
phase wandering. This effect is unlikely to affect a semicoherent search like the one here reported, 
but it could have a significant impact during the follow-up stage, where longer coherence times are 
used \cite{Mukherjee:2017qme, Ashton:2017wui}.}. 
Regions excluded by the spindown limit correspond to shaded areas in Fig.~\ref{fig:reach}.

Equatorial ellipticity values can be constrained below $\epsilon=10^{-5}$ for sources in binary systems such as 
the ones analyzed by this search located at $1\;\textrm{kpc}$ emitting within the $150 - 300\;\textrm{Hz}$ band.
Constraints below $\epsilon=10^{-4}$ can be set for sources located at $2\;\textrm{kpc}$ emitting within the
$75 - 150\;\textrm{Hz}$ band. These sensitivities approach the expected allowed maximum ellipticities of 
relativisitic stars, which range from the order of $10^{-6} - 10^{-7}$ to values around  $10^{-5}$ for more 
exotic equations of state \cite{JohnsonMcDaniel:2012wg}.

Future enhancements of the terrestrial gravitational wave detector network will improve our sensitivity to
fainter gravitational wave signals, providing a valuable tool to prospect the expected population of galactic 
NSs in binary systems \cite{KAGRA, Maggiore:2019uih, Reitze:2019iox, Miller:2014kma,Aasi:2013wya}.

\section{Acknowledgements}
The authors gratefully acknowledge the support of the United States
National Science Foundation (NSF) for the construction and operation of the
LIGO Laboratory and Advanced LIGO as well as the Science and Technology Facilities Council (STFC) of the
United Kingdom, the Max-Planck-Society (MPS), and the State of
Niedersachsen/Germany for support of the construction of Advanced LIGO 
and construction and operation of the GEO600 detector. 
Additional support for Advanced LIGO was provided by the Australian Research Council.
The authors gratefully acknowledge the Italian Istituto Nazionale di Fisica Nucleare (INFN),  
the French Centre National de la Recherche Scientifique (CNRS) and
the Netherlands Organization for Scientific Research, 
for the construction and operation of the Virgo detector
and the creation and support  of the EGO consortium. 
The authors also gratefully acknowledge research support from these agencies as well as by 
the Council of Scientific and Industrial Research of India, 
the Department of Science and Technology, India,
the Science \& Engineering Research Board (SERB), India,
the Ministry of Human Resource Development, India,
the Spanish Agencia Estatal de Investigaci\'on,
the Vicepresid\`encia i Conselleria d'Innovaci\'o, Recerca i Turisme, the Conselleria d'Educaci\'o i Universitat and the Direcci\'o General de Pol\'itica Universitaria i Recerca del Govern de les Illes Balears,
the Conselleria d'Innovaci\'o, Universitats, Ci\`encia i Societat Digital de la Generalitat Valenciana and
the CERCA Programme Generalitat de Catalunya, 
the Barcelona Supercomputing Center - Centro Nacional de Supercomputaci\'on, Spain,
the National Science Centre of Poland and the Foundation for Polish Science (FNP),
the Swiss National Science Foundation (SNSF),
the Russian Foundation for Basic Research, 
the Russian Science Foundation,
the European Commission,
the European Regional Development Funds (ERDF),
the Royal Society, 
the Scottish Funding Council, 
the Scottish Universities Physics Alliance, 
the Hungarian Scientific Research Fund (OTKA),
the French Lyon Institute of Origins (LIO),
the Belgian Fonds de la Recherche Scientifique (FRS-FNRS), 
Actions de Recherche Concert\'{e}es (ARC) and
Fonds Wetenschappelijk Onderzoek - Vlaanderen (FWO), Belgium,
the Paris \^{I}le-de-France Region, 
the National Research, Development and Innovation Office Hungary (NKFIH), 
the National Research Foundation of Korea,
the Natural Science and Engineering Research Council Canada,
Canadian Foundation for Innovation (CFI),
the Brazilian Ministry of Science, Technology, and Innovations,
the International Center for Theoretical Physics South American Institute for Fundamental Research (ICTP-SAIFR), 
the Research Grants Council of Hong Kong,
the National Natural Science Foundation of China (NSFC),
the Leverhulme Trust, 
the Research Corporation, 
the Ministry of Science and Technology (MOST), Taiwan,
the United States Department of Energy,
and
the Kavli Foundation.
The authors gratefully acknowledge the support of the NSF, STFC, INFN and CNRS for provision of computational resources.
This paper has been assigned document number LIGO-P2000298.

\bibliographystyle{apsrev4-2}
\bibliography{references} 

\appendix
\section{Frequency bands containing outliers}
\label{appendix}
We provide a list of frequency bands in which outliers surviving the follow up were found. These outliers were discarded
due to their inconsistent behavior with respect to an astrophysical signal, as discussed in Sec.~\ref{sec:fu}.
Table~\ref{ref:spiky_outliers} lists frequency bands where line-crossing outliers were found.
Table~\ref{ref:persistent_outliers} corresponds to frequency bands presenting outliers discarded by the detector
consistency veto.
In both tables, overlapping frequency bands are merged together for the sake of compactness.

\begin{table*}
    \begin{tabular}{ccccccc}
\toprule
    Min. frequency [Hz] &  Max. frequency [Hz] &  First timestamp [GPS] &  Last timestamp [GPS] &  Duration [days] & Detector & Listed \\
\midrule
              55.605 &               55.606 &             1239194624 &            1253831680 &              169 &             H1 & Yes \\
              57.588 &               57.592 &             1243670016 &            1253770240 &              116 &             H1 & Yes \\
              59.501 &               59.513 &             1238290944 &            1253831680 &              179 &             H1 & Yes \\
              62.474 &               62.478 &             1244523008 &            1251895296 &               85 &             H1 & Yes \\
              64.257 &               64.266 &             1241190400 &            1252681728 &              133 &             H1 & No \\
              64.284 &               64.291 &             1238955520 &            1253831680 &              172 &             H1 & Yes \\
              64.470 &               64.475 &             1239557632 &            1253770240 &              164 &             H1 & Yes \\
              64.403 &               64.408 &             1240341504 &            1253588992 &              153 &             H1 & Yes \\
              64.364 &               64.375 &             1238529536 &            1253831680 &              177 &             H1 & Yes \\
              64.415 &               64.417 &             1239375872 &            1253831680 &              167 &             H1 & Yes \\
              70.124 &               70.128 &             1238408192 &            1245484544 &               81 &             H1 & Yes \\
              80.067 &               80.070 &             1238351360 &            1238351360 &                0 &             H1 & No \\
              83.307 &               83.316 &             1238351360 &            1253831680 &              179 &             H1 & Yes \\
              83.446 &               83.448 &             1238831616 &            1253831680 &              173 &             H1 & Yes \\
              85.712 &               85.715 &             1239194624 &            1253831680 &              169 &             H1 & Yes \\
              85.964 &               85.965 &             1239738880 &            1253165568 &              155 &             H1 & Yes \\
              99.966 &               99.979 &             1238290944 &            1253831680 &              179 &             H1 & Yes \\
             107.113 &              107.119 &             1241129984 &            1253709824 &              145 &             H1 & Yes \\
             140.253 &              140.254 &             1238955520 &            1245424128 &               74 &             H1 & Yes \\
             151.800 &              151.800 &             1253105152 &            1253105152 &                0 &             H1 & Yes \\
             199.946 &              199.955 &             1242149888 &            1253770240 &              134 &             H1 & Yes \\
             213.301 &              213.301 &             1242339840 &            1242339840 &                0 &             L1 & Yes \\
\bottomrule
\end{tabular}

    \caption{Frequency bands containing line-crossing outliers. As described in Sec.~\ref{sec:transient_lines},
    these could be produced because of the presence of a transient instrumental
    artifact or the frequency evolution of a candidate drifting away from the spectral disturbance. 
    Overlapping frequency bands were grouped together for the sake of simplicity. 
    Outliers belonging to this category show at least one per-segment $\mathcal{F}$-statistic value greater than
    $100$.
    Timestamps refer to the first and last coherent segments ($T_\textrm{coh} \simeq 17\textrm{h}$) for which at 
    least one of those candidates showed an $\mathcal{F}$-statistic value over 50.
    The last column relates these bands to the list of unidentified lines \cite{UnidentifiedLines}.
    }
    \label{ref:spiky_outliers}
\end{table*}

\begin{table*}
    \begin{tabular}{ccc}
\toprule
 Min. frequency [Hz] &  Max. frequency [Hz] & Listed \\
\midrule
              53.709 &               53.721 &                      No \\
              55.603 &               55.609 &                     Yes \\
              57.583 &               57.600 &                     Yes \\
              62.823 &               62.828 &                     Yes \\
              64.400 &               64.411 &                     Yes \\
              83.442 &               83.453 &                     Yes \\
              85.815 &               85.824 &                     No \\
\bottomrule
\end{tabular}

    \caption{Frequency bands containing outliers discarded by the detector consistency veto as described 
    in Sec.~\ref{sec:consistency_veto}. Overlapping frequency bands were grouped together for the sake
    of simplicity. The last column relates these bands to the list of unidentified lines of the H1 detector
    \cite{UnidentifiedLines}.}
    \label{ref:persistent_outliers}
\end{table*}

\clearpage

\iftoggle{endauthorlist}{
 %
 %
 \let\author\myauthor
 \let\affiliation\myaffiliation
 \let\maketitle\mymaketitle
 \pacs{}

 \newpage
 \maketitle
}

\end{document}